\documentclass[showpacs,prd,twocolumn,floatfix,
nofootinbib,superscriptaddress]{revtex4}
\usepackage{graphicx}
\usepackage{color}
\usepackage{verbatim}

\definecolor{Blue}{rgb}{0.,0.,1.}
\definecolor{Green}{rgb}{0.,1.,0.}
\definecolor{Red}{rgb}{1.,0.,0.}

\newcommand{\be}{\begin{equation}}
\newcommand{\ee}{\end{equation}}
\newcommand{\nl}{\nonumber \\}

\newcommand{\psibar}{\overline{\Psi}}

\begin{document}


\title{ $D \rightarrow K, l \nu$ Semileptonic Decay Scalar Form Factor 
and $|V_{cs}|$ from Lattice QCD}

\author{Heechang Na}
\affiliation{Department of Physics,
The Ohio State University, Columbus, OH 43210, USA}
\author{Christine T.\ H.\ Davies}
\affiliation{Department of Physics \& Astronomy,
University of Glasgow, Glasgow, G12 8QQ, UK}
\author{Eduardo Follana}
\affiliation{Departmento de Fisica Teorica, Universidad de Zaragoza, Zaragoza, Spain}
\author{G.\ Peter Lepage}
\affiliation{Laboratory of Elementary Particle Physics,
Cornell University, Ithaca, NY 14853, USA}
\author{Junko Shigemitsu}
\affiliation{Department of Physics,
The Ohio State University, Columbus, OH 43210, USA}

\collaboration{HPQCD Collaboration}
\noaffiliation
\date{\today  }


\begin{abstract}
We present a new study of D semileptonic decays on the lattice 
which employs the Highly Improved Staggered Quark (HISQ) action for 
both the charm and the light valence quarks. 
We work with MILC unquenched $N_f = 2 + 1$ 
lattices and determine the scalar form factor $f_0(q^2)$ for $D \rightarrow K, l \nu$ 
semileptonic decays. 
The form factor is obtained from a scalar current matrix element that does not require any operator matching.
We develop a new approach to carrying out chiral/continuum 
extrapolations of $f_0(q^2)$.  The method uses the kinematic ``$z$'' variable 
instead of $q^2$ or the kaon energy $E_K$ and is applicable over the entire 
physical $q^2$ range.  
 We find $f^{D \rightarrow K}_0(0) \equiv f^{D \rightarrow K}_+(0) = 0.747(19)$ 
in the chiral plus 
continuum limit and hereby improve the theory error on this quantity by a 
 factor of $\sim$4 compared to previous lattice determinations. Combining the new
 theory result with  recent experimental
 measurements of the product $f^{D \rightarrow K}_+(0) * |V_{cs}| $
from BaBar and CLEO-c leads to a very precise 
direct determination of the CKM matrix element  
$|V_{cs}| $, 
$|V_{cs}| = 0.961(11)(24)$, where the first error comes from experiment and the second 
is the lattice QCD theory error.
We calculate the ratio $f^{D \rightarrow K}_+(0)/f_{D_s}$ and find $2.986 \pm 0.087$ GeV$^{-1}$ and show that this agrees with experiment.

\end{abstract}


\maketitle


\section{Introduction}
Independent determinations of each of the Cabibbo-Kobayashi-Maskawa (CKM) 
matrix elements and checks of three generation unitarity provide 
stringent consistency tests of the Standard Model and have become 
an important part of flavor physics. For instance, first row unitarity 
has now been checked to very high accuracy with $|V_{ud}|^2 + |V_{us}|^2 
+ |V_{ub}|^2 - 1 = - 0.0001(6)$ \cite{flavianet}. 
 Such precision became possible when both  
experiment and lattice QCD theory inputs to the  determination of $|V_{us}|$ 
 reached subpercent levels of accuracy (the contribution from $|V_{ub}|$ to 
the unitarity sum is negligible and $|V_{ud}|$ is known with $\sim 0.02$\%
 errors).

\vspace{.05in}
In contrast to the elements of the 1st row, 
direct determinations of the 2nd row matrix elements are still much 
less precise.  
The latest PDG summary \cite{pdg} quotes 4.8\%, 3.5\% and 3.2\% errors for 
$|V_{cd}|$, $|V_{cs}|$ and $|V_{cb}|$ respectively when one considers all 
ways of extracting the matrix elements. If one focuses just on 
 determinations of $|V_{cd}|$ and $|V_{cs}|$ from semileptonic $D \rightarrow 
\pi, l \nu$ and $D \rightarrow K, l \nu$ decays then until recently 
the total error has been at the 10\% level and dominated by lattice QCD theory errors.
Furthermore tests of 2nd row unitarity stands at 
$|V_{cd}|^2 + |V_{cs}|^2 + |V_{cb}|^2 = 1.101 \pm 0.074$ and for the 2nd column at
$|V_{us}|^2 + |V_{cs}|^2 + |V_{ts}|^2 = 1.099 \pm 0.074$ \cite{pdg}. 
 In both cases the error 
is dominated by the uncertainty in $|V_{cs}|$. Clearly reducing the errors in 
$|V_{cs}|$ will have immediate and significant impact on flavor physics and on CKM 
unitarity tests.  On the experimental front Belle \cite{belle}, 
BaBar \cite{babar} and CLEO-c \cite{cleo} have all 
recently published precise measurements of the combination 
$f^{D \rightarrow K}_+(0) * |V_{cs}|$ with 3.3\%, 1.4\% and 1.1\% errors 
respectively.  These precise measurements can be turned into accurate 
$|V_{cs}|$ determinations if and only 
if theory can provide the form factor $f^{D \rightarrow K}_+(0)$ 
with comparable precision.  The main goal of the current work was to improve
lattice QCD calculations of $f^{D \rightarrow K}_+(0)$ and the outcome is that 
we   have now succeeded in reducing 
the theory errors from 10\% down to 2.5\%.  Innovations that made this
dramatic improvement in errors possible 
include the employment of a new and better action for charm quarks,  the use of
an absolutely normalized hadronic matrix element that does not require any operator matching, 
 improved analysis tools  
for lattice data and a new method for carrying out chiral/continuum extrapolations.

\vspace{.05in}
One reason for the large disparity in the size of lattice QCD errors between 
kaon and $D$ meson systems in the past 
has been the challenge of simulating quarks with masses 
as large as that of the charm quark.  If ``$a$'' is the 
lattice spacing and hence $\sim 1/a $ the cutoff in the theory, 
then for charm quarks it 
was thought to be difficult to satisfy $a m_c = m_c/{\rm cutoff} \ll 1$.  
In the past this problem was circumvented by employing effective 
theories to handle the charm quark,  nonrelativistic (NR)QCD, HQET 
or the ``heavy clover'' action of the Fermilab lattice collaboration. 
The first $N_f = 2+1$ unquenched studies of $D$ semileptonic decays on the lattice 
were carried out by the Fermilab Lattice and MILC collaborations using an effective theory 
for charm \cite{fermimilc05}. 
  This pioneering work 
predicted the shape of the form factors as a function of $q^2$ prior to 
subsequent verification by experiment. The theory errors, however, were quite large 
at $\sim 10$\%. The calculations by 
the Fermilab Lattice and MILC collaborations are being improved upon 
and their theory errors should be reduced significantly soon \cite{fermimilc10}.
 On the other hand, 
working with  effective theories 
typically leads to larger statistical and systematic errors than when employing 
``relativistic'' quark actions as can be done for kaon physics. 
This includes uncertainties coming from matching of heavy-light currents 
responsible for heavy meson leptonic or semileptonic decays, and from the tuning of 
heavy quark masses, all procedures that are much more complicated in 
effective theories.

\vspace{.05in}
In recent years there has been a major shift in lattice 
simulations with charm quarks.  With the advent of highly improved 
lattice quark actions the concerns described above of $m_c/{\rm cutoff}$ 
being too large have been overcome and several 
collaborations are now working on charm physics without resorting to 
effective theories. 
 In 2007 the HPQCD collaboration introduced the ``Highly Improved Staggered
 Quark'' (HISQ) action \cite{hisq}. 
Many lattice artifacts, including all ${\cal O}((am_c)^2)$
 discretization effects and all ${\cal O}(\alpha_s (am_c)^2)$ and ${\cal O}
 ((am_c)^4)$ effects at leading order in the charm quark velocity $\frac{v}{c}$
 have been removed.  It then becomes feasible to simulate charm quarks
on the lattice  in a fully relativistic
 setting without introducing large discretization errors as long as
 one works with lattice spacings $a \leq \sim 0.15$fm. This last condition is easily
 met nowadays in typical lattice simulations.
Another approach to charmed meson leptonic and semileptonic 
decays  using relativistic charm quarks is 
being pursued by the European Twisted Mass (ETM) collaboration \cite{etmc}.
They employ a special version of Wilson type quark action called 
the ``Twisted Mass'' quark action where discretization errors come in at 
${\cal O}(a^2)$.

\vspace{.05in}
The HPQCD collaboration has successfully applied HISQ charm quarks to studies 
of $D$ and $D_s$ meson leptonic decays \cite{hpqcdfds}. 
 This formalism significantly reduced 
lattice errors in decay constant calculations.  We have now also initiated 
$D$ meson semileptonic decay studies based on HISQ charm and light valence 
quarks. In reference \cite{nalat09} we developed and  tested our approach on a simpler test 
case of a fictitious $D_s \rightarrow \eta_s, l \nu$ semileptonic decay 
($\eta_s$ refers to a  pseudoscalar $s - \overline{s}$ bound state).
In the current paper we present the first application of HISQ charm quarks to realistic 
$D$ meson semileptonic decays.  More specifically we calculate the 
scalar form factor $f_0(q^2)$ for $D \rightarrow K, l \nu$ decays.   
As mentioned  above already 
experiment provides us with the product $f_+(0) * |V_{cs}|$.
 Then using the kinematic relation $f_0(0) = f_+(0)$, an accurate calculation of $f_0(q^2)$ 
from the lattice leads to a precise determination of the CKM matrix element 
$|V_{cs}|$.

\vspace{.05in}
We have carried out simulations on three of the MILC ``coarse'' ensembles 
with lattice spacing $a \sim 0.12$fm and two of the ``fine'' ensembles with 
$a \sim 0.09$fm.  Details of these ensembles are listed in Table~\ref{T.lat}. The five 
ensembles provide enough variation and information to allow sensible 
chiral and continuum extrapolations to the physical world. 

\vspace{.1in}
\begin{table}
\caption{ Details of MILC configurations employed in this article.
 $N_{tsrc}$ is the number
of time sources used per configuration.
 All sea quark masses are given in the
MILC collaboration normalization convention with
 $u^{plaq}_0 = \langle plaquette \rangle^{1/4}$. Values for 
the scale variable $r_1$ in lattice units, $r_1/a$, are 
taken from Ref. \cite{milc}.  Errors in this ratio are 
 at the $\sim$0.1\% level.
}
\label{T.lat}
\begin{center}
\begin{tabular}{|c|c|c|c|c|c|c|}
\hline
Set &$r_1/a$   & $a u_0 m_{sea}$ &  $ u^{plaq}_0$ & $N_{conf}$&
$N_{tsrc}$  & $L^3 \times N_t$ \\
\hline
\hline
C1  & 2.647 & 0.005/0.050 & 0.8678   & 600 & 2 & $24^3 \times 64$ \\
C2  & 2.618 & 0.010/0.050 & 0.8677  & 600 & 2 & $20^3 \times 64$ \\
C3  & 2.644 & 0.020/0.050 & 0.8688   & 600 & 2 & $20^3 \times 64$ \\
\hline
F1  & 3.699 & 0.0062/0.031 & 0.8782 & 600 & 4  & $28^3 \times 96$ \\
F2  & 3.712 & 0.0124/0.031 & 0.8788 & 600 & 4 & $28^3 \times 96$ \\
\hline
\end{tabular}
\end{center}
\end{table}

\vspace{.05in}
In the next section we review the formalism for extracting semileptonic decay 
form factors from hadronic matrix elements of appropriate heavy-light 
currents.  We describe the advantages of working with the same 
relativistic action for both the heavy and the light quarks and explain 
why the form factor at $q^2 = 0$ is most accurately extracted from hadronic 
matrix elements of the scalar current as opposed to from the vector current.
 In section III we introduce the 
HISQ action and describe how action parameters such as bare quark masses 
were tuned. Section IV provides further details of our simulations including 
benefits derived from using ``random wall'' sources, in particular when 
simulating kaons with nonzero momenta.
Section V describes our fitting strategy.  We have invested considerable effort 
into developing improved fitting methods in order to extract the form factors 
of interest with subpercent errors for each ensemble and for all kaon momenta
needed to cover the physical $q^2$ range.

\vspace{.05in}
  In section VI we 
take the form factor results from the five ensembles and extrapolate to 
the chiral/continuum limit.  To this end we have developed a new extrapolation 
method that can be used over the entire physical $q^2$ range, $(M_D-M_K)^2 
\geq q^2 \geq 0$.  Since we are interested in the form factor at $q^2 = 0$, 
it is important that any extrapolation scheme work all the way down to 
$q^2 = 0$ where the kaon in the $D$ meson rest frame has energy $E_K \approx 
1 \rm{GeV}$. The new approach uses the ``$z$-expansion'' of Refs. 
\cite{boyd, arnesen, hill} to 
parameterize the kinematics (the dependence on $E_k$ or equivalently on 
$q^2$).  The coefficients of this expansion are then allowed to be functions 
of the light and strange quark masses and of the lattice spacing.  We find that 
good fits to all our data are possible with such an ansatz and the resulting 
$f_0(0)$ in the chiral/continuum limit is very stable against higher order 
terms in this  ansatz.

\vspace{.05in}
Section VII summarises our final results for $f_0(0) = f_+(0)$
at the physical point and explains our 
error budget.  We incorporate experimental input from BaBar \cite{babar} and CLEO-c
\cite{cleo}
to determine the CKM matrix element $|V_{cs}|$ and compare with values
listed in the PDG and with expectations from CKM unitarity.
In Section VIII we collect results obtained as ``side products'' of
our analysis of semileptonic decay three-point hadronic matrix elements,
namely results coming from two-point correlators such as
decay constants of the pion, kaon, $D$ and $D_s$ mesons. These two-point
results provide nontrivial tests of our mass tunings, fitting and
chiral/continuum extrapolation strategies leading to greater confidence
in our form factor (i.e. three-point) results as well.
Finally section IX gives a summary and addresses future plans.
Several appendices cover details of Bayesian fits employed in this article.
And in Appendix C we carry out the chiral/continuum extrapolation of $f_0(q^2)$ using
an approach that differs completely from the $z$-expansion method of
section VI and is based instead on Chiral Perturbation theory.  We show that the two
extrapolation methods give results in very good agreement with each other.

\section{Formalism}
To study the process $D \rightarrow K, l \nu$ one needs to evaluate the 
matrix element of the charged electroweak current between the $D$ and the $K$ meson 
states, $\langle K | (V^\mu - A^\mu) | D \rangle $.  Only the 
vector current $V^\mu$ contributes to the pseudoscalar-to-pseudoscalar 
amplitude and the matrix element can be written in terms of two 
form factors $f_+(q^2)$ and $f_0(q^2)$, where $q^\mu = p^\mu_D - p^\mu_K$ 
is the four-momentum of the emitted W-boson.
\begin{eqnarray}
\label{vmu}
\langle K| V^\mu | D \rangle &=&  f^{D \rightarrow K}_+(q^2) \left [ p^\mu_D + p^\mu_K
- \frac{M^2_D - M^2_K}{q^2} \, q^\mu \right ] \nl
 &+&  f^{D \rightarrow K}_0(q^2) \frac{M_D^2 - M_K^2}{q^2} q^\mu 
\end{eqnarray}
with $V^\mu \equiv \psibar_s \gamma^\mu \Psi_c$. 
As described below, we find it useful to consider also the matrix element of 
the scalar current $S \equiv \psibar_s \Psi_c$,
\be
\label{scalar}
\langle K| S | D \rangle = \frac{M_D^2 - M_K^2}{m_{0c} - m_{0s}} f^{D \rightarrow K}_0(q^2).
\ee
In continuum QCD one has the PCVC (partially conserved vector current)
relation and the vector and scalar currents obey,
\be
\label{pcvc}
q^\mu \langle V^{cont.}_\mu \rangle = (m_{0c} - m_{0s}) \langle S^{cont.} \rangle.
\ee
In fact PCVC is the reason why the same form factor $f^{D \rightarrow K}_0(q^2)$  appears 
in eqs.(\ref{vmu}) and (\ref{scalar}).  On the lattice it is often 
much more convenient to simulate with vector currents $\psibar_{Q1} \gamma^\mu 
\Psi_{Q2}$ that are not exactly conserved at finite lattice spacings even 
for $Q1 = Q2$.
  Such non-exactly-conserved 
currents need to be renormalized and acquire Z-factors.  We are able to carry out 
fully nonperturbative renormalization of the lattice vector current by 
imposing PCVC.  In the $D$ meson rest frame the condition becomes,
\be
\label{latpcvc}
(M_D - E_K) \langle V_0^{latt.} \rangle Z_t +
\vec{p}_K \cdot \langle \vec{V}^{latt.} \rangle Z_s =
(m_{0c} - m_{0s}) \langle S^{latt.} \rangle.
\ee

\vspace{.05in}
We have checked the feasibility of this renormalization scheme and 
extracted preliminary 
 $Z_t$ and $Z_s$ values for the test case of $D_s \rightarrow \eta_s, l \nu$ 
mentioned above in Ref.\cite{nalat09}.  We plan to apply this fully nonperturbative 
renormalization scheme to evaluate $\langle \pi | V^\mu | D \rangle$ and 
$\langle K | V^\mu | D \rangle$ relevant for realistic $D \rightarrow \pi, l \nu$ 
and $D \rightarrow K, l \nu$ semileptonic decays in the near future.  In the present 
article, however, we will focus on the form factor $f_+(q^2)$ just at 
$q^2 = 0$, since this is all that is needed to extract $|V_{cs}|$.  
We do this by exploiting  the kinematic identity $f_+(0) = f_0(0)$, and 
concentrating on determining the scalar form factor $f_0(q^2)$ as accurately as 
possible.  The best way to proceed is to evaluate the hadronic matrix 
element of the scalar current rather than of the vector current.  
From eq.(\ref{scalar}) one then has, 
\be
\label{f0}
f^{D \rightarrow K}_0(q^2) =
 \frac{(m_{0c} - m_{0s}) \langle K | S | D \rangle}{M_D^2 - M_K^2}.
\ee
The numerator on the right-hand-side is a renormalization 
group invariant combination.
This is true even in our lattice formulation, because we use the same relativistic action for both the heavy and the light valence quarks.
Moreover, eq.~(\ref{f0}) allows 
a lattice determination of $f_0(q^2)$ and hence also of $f_+(0) = f_0(0)$ 
without any need for operator 
matching.  
Using eq.~(\ref{f0}) and going to the continuum limit is straightforward, because our action is so highly improved even for heavy quarks.

\section{The HISQ Action and Tuning of Action Parameters}

\vspace{.1in}
\begin{table}
\caption{ Action parameters and $\eta_c$ and $\eta_s$ masses.  $1 + \epsilon$ 
enters the coefficient of the Naik term for charm propagators.  
}
\label{T.qmass}
\begin{center}
\begin{tabular}{|c|c|c|c|c|c|c|}
\hline
Set &$a m_{0c}$   & $a m_{0s}$ &  $ a m_{0l}$ & $ 1 + \epsilon$&
$aM_{\eta_c}$  & $aM_{\eta_s}$ \\
\hline
\hline
C1  & 0.6207 & 0.0489 & 0.0070 & 0.780 &1.7887(1) &0.4111(2) \\
C2  & 0.6300 & 0.0492 & 0.0123 & 0.774 &1.8085(1) &0.4143(2) \\
C3  & 0.6235 & 0.0491 & 0.0246 & 0.778 &1.7907(1) &0.4118(2) \\
\hline
F1  & 0.4130 & 0.0337 & 0.00674 & 0.893 &1.2807(1) &0.2942(1) \\
F2  & 0.4120 & 0.0336 & 0.01350 & 0.894 &1.2751(1) &0.2931(2) \\
\hline
\end{tabular}
\end{center}
\end{table}

\begin{figure}
\includegraphics*[width=7.0cm,height=8.0cm,angle=-90]{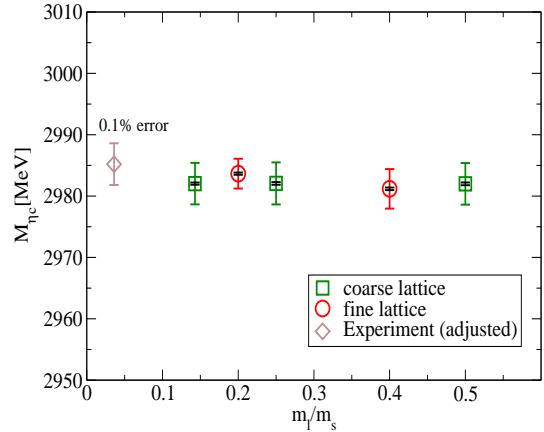}
\caption{
 Tuning of the charm quark mass via the $\eta_c$ meson mass. Errors on 
the simulation results include statistical plus the $\sim 0.1$\% 
errors coming from $r_1/a$. The smaller black error bars on the lattice data indicate the statistical errors. The ``experimental'' $\eta_c$ mass has 
been adjusted to take into account the lack of annihilation and 
electromagnetic effects in our lattice calculation.
 }
\label{etac}
\end{figure}

\begin{figure}
\includegraphics*[width=7.0cm,height=8.0cm,angle=-90]{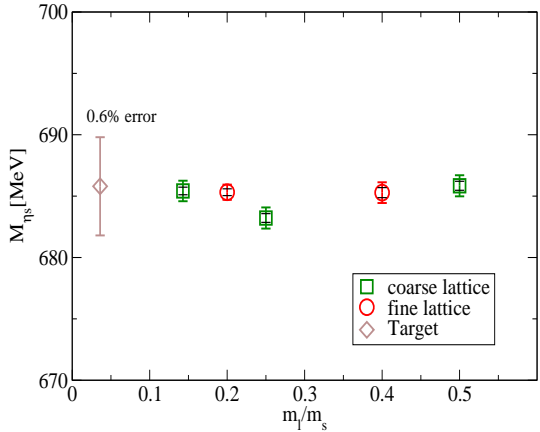}
\caption{
Tuning of the strange quark mass via the $\eta_s$
 }
\label{etas}
\end{figure}

The HISQ action was introduced in Ref.\cite{hisq} and represents the next level of 
improvement of staggered quarks beyond the AsqTad \cite{asqtad}
 action.  Relative to the latter, 
the HISQ action reduces taste breaking effects by approximately a factor of three 
on MILC coarse and fine lattices. For heavy quarks such as charm the HISQ action 
includes one adjustable parameter $\epsilon$ which modifies the ``Naik'' term 
already present in the AsqTad action $\frac{1}{6} a^2\Delta^3_\mu \rightarrow 
\frac{1 + \epsilon}{6} a^2\Delta^3_\mu$.  In Ref.\cite{hisq} $\epsilon$ was adjusted 
nonperturbatively to get the correct dispersion relation for the 
$\eta_c$ meson.  It was found that one ends up with $\epsilon$'s close to  
estimates coming from requiring that the tree-level quark propagator have 
speed of light $c(p) = 1$. In the current simulations we have 
set $\epsilon$ in the charm propagators equal to its tree-level value. 
The size of $\epsilon$ decreases rapidly with decreasing $am_q$, so for strange and 
light quarks $\epsilon$ can safely be set equal to zero.
Table~\ref{T.qmass} lists the bare valence masses and the $\epsilon(charm)$ values employed for 
the five ensembles. 
The bare charm and strange quark masses were tuned in order to obtain 
the correct $\eta_c$ and $\eta_s$ meson masses.  $a m_{0l}$ was chosen so that 
the ratio of HISQ valence masses 
$m_{0l}(valence)/m_{0s}(valence)$ ended up close to the analogous ratio of 
light and ``physical'' strange quark masses for AsqTad quarks on the same ensemble~\footnote[1]{Note that HISQ and AsqTad masses are not numerically the same, since the two actions are different.}. 
This makes the valence light quark mass  approximately equal to the sea 
light quark mass, both measured relative to a physical, tuned strange quark mass.
 Figures \ref{etac} and \ref{etas} show our results for the $\eta_c$ and $\eta_s$ masses 
on the five ensembles together with the target values.
  One sees that tuning has been achieved very accurately.  For $\eta_c$ our target 
value is $M_{\eta_c}^{target} = 2.9852(34)$GeV which differs slightly from the 
experimental $M_{\eta_c}^{exper.} = 2.9803$GeV since we adjust for 
the absence of electromagnetic and annihilation effects in the lattice simulations
 \cite{r1}.
The target value for the $\eta_s$ is $M_{\eta_s} = 0.6858(40)$GeV~\cite{r1}.  
The data points in Figs.~\ref{etac} and \ref{etas} show only statistical and $r_1/a$ errors.
 We first determine 
$r_1 \times M_{meson}$  using the precisely know $r_1/a$ for each ensemble. 
At this point our data points have errors coming from both statistics and 
from the $\sim$0.1\% uncertainty in $r_1/a$ (with the latter dominating). 
For the purposes of using a physical scale on the vertical axis we then 
convert $r_1 \times M_{meson}$ to MeV using $r1 = 0.3133(23)$fm ($r_1^{-1} = 0.6297(46)$GeV)
 \cite{r1}.  
For reasons explained below, we find it more informative not to include the 
$\sim 0.7$\% uncertainty in the physical value of 
$r_1$ in these plots.  Including this error will affect all five data points 
in the same way without changing relative uncertainties.

\begin{figure}
\includegraphics*[width=7.0cm,height=8.0cm,angle=-90]{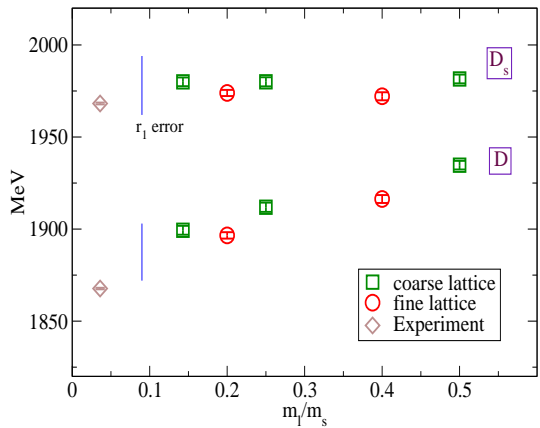}
\caption{
The $D$ and $D_s$ meson masses.  As explained in the text although the 
vertical axis uses physical units, the data points show 
statistical and $r_1/a$ errors only.  The experimental points have 
not been adjusted for electromagnetic effects that are absent in the 
lattice simulations. Such effects have been estimated to be less 
than 0.2\%~\cite{fds2}.
 }
\label{DDs}
\end{figure}

\vspace{.05in}
Once the bare charm and strange quark masses have been fixed for each ensemble 
there are no adjustable parameters left when one goes on to determining other 
meson masses such as $M_D$ or $M_{D_s}$, decay constants $f_D$, $f_{D_s}$, $f_K$ etc. 
or semileptonic form factors.  In Fig.~\ref{DDs} we show results for $M_D$ and $M_{D_s}$.  
Again the errors on the  data points reflect statistical and $r_1/a$ errors only.  
Any changes in the physical value of $r_1$ will shift all five data points 
uniformly without affecting their relative positions.  Furthermore 
chiral/continuum extrapolations would be carried out on 
$r_1 \times M_{meson}$ with $r_1$ and its error coming in only after having extracted the 
physical limit. By omitting the full $r_1$ errors in 
Fig.~\ref{DDs} one can more easily identify discretization effects and light quark 
mass dependence. For instance, for $M_{D_s}$ the difference between the coarse and 
fine ensemble results is at the $\sim6$MeV level or $\sim 0.3$\% and the sea light 
quark mass dependence is essentially nonexistent.  The 0.3\% discretization effect 
should be compared to the $\sim 0.7$\% uncertainty in $r_1$.
One lesson to be learnt from this is the importance of tuning quark masses accurately 
enough so that results on the different ensembles agree to within the smaller 
$r_1/a$ errors and not just to within the larger $r_1$ error.  Otherwise it would 
not be possible to have data points lying along smooth curves as in Fig.~\ref{DDs} where 
discretization and light quark mass dependence can be clearly 
identified and distinguished from mass tuning and $r_1$ errors.
  Based on Figs.~\ref{etac}, \ref{etas} and \ref{DDs} we believe 
the HISQ action parameters have been fixed accurately enough in preparation for 
going on to D semileptonic decays.  Further consistency checks, such as 
determinations of decay constants will be given in section VIII.

\section{Simulation Details}
The goal is to determine the hadronic matrix element $\langle K | S | D \rangle$ 
in eq.(\ref{f0}) via numerical simulations.  The starting point is the 
three-point correlator,
\begin{eqnarray}
\label{thrpnt}
& & C^{3pnt}(t_0,t,T,\vec{p}_K) = 
\frac{1}{L^3} \sum_{\vec{x}}
\sum_{\vec{y}} \sum_{\vec{z}} e^{i\vec{p}_K \cdot (\vec{z} - \vec{x})} \nl
&& \qquad \langle \Phi_K(\vec{x},t_0) \,\tilde{S}(\vec{z},t) \, 
\Phi^\dagger_D(\vec{y},t_0-T)
 \rangle. 
\end{eqnarray}
$\Phi^\dagger_D$ and $\Phi_K$ are interpolating operators that create 
a $D$ meson or annihilate a kaon respectively and $\tilde{S} \equiv a^3 S$ is 
the scalar current in lattice units.  We also work with dimensionless 
fermion fields. 
 Eq.~(\ref{thrpnt}) 
corresponds to first creating a zero momentum $D$ meson  at time 
$t_0 - T$ which propagates to time slice t with $t_0 \geq t \geq t_0 - T$. 
At time slice $t$ the scalar current $\tilde{S} = \psibar_s \Psi_c$ 
converts the charm quark inside the $D$ meson into a strange quark 
while inserting momentum $\vec{p}_K$.  The resulting kaon is then annihilated 
at time $t_0$.  

\vspace{.05in}
In our simulations we have picked $t_0$ the location of the kaon 
operator $\Phi_K$ randomly and differently for each configuration 
in order to reduce autocorrelations. We also worked with $N_{tsrc}$ values 
of $t_0$ 
per configuration (see Table~\ref{T.lat}) placed $N_t/N_{tsrc}$ time slices apart 
($N_t$ is the total number of time slices for our lattices).
Once $t_0$ was fixed, for each configuration 
we obtained results for several $T$ values, $T =  15$ and $16$ on coarse 
and $T = 19$, $20$, and $23$ on fine lattices.  We will see later that having 
data at many $T$ values significantly reduces errors in extracted 
three-point amplitudes.  To further improve statistics, for each 
 $t_0$ and $T$ value we also evaluated the time-reversed three-point 
correlator, essentially the same as eq.(\ref{thrpnt}) but with 
$\Phi^\dagger_D$ acting on time slice $t_0 + T$ and the scalar current 
inserted at $T+ t_0 \geq t \geq t_0$.

\vspace{.05in}
As is well known, with the HISQ action, each flavor of quark comes in $N_{taste}$ 
copies called ``tastes''.  One has $N_{taste} = 4$ when working with one-component 
staggered fields and $N_{taste} = 16$ for four-component ``naive'' fields.
In the naive fields language the interpolating operators $\Phi_{D/K}$ and 
the scalar current $\tilde{S}$ become,
\be
\Phi^\dagger_D = \frac{1}{4}\psibar_c \gamma_5 \Psi_l, \qquad 
 \Phi_K = \frac{1}{4} \psibar_l \gamma_5 \Psi_s,
\ee
and
\be
\label{scalarc}
\tilde{S} = \psibar_s \Psi_c.
\ee
These are all single site bilinears. $\Phi_{D/K}$ correspond to 
taste non-singlet ``Goldstone'' pseudoscalars and the factors of 
$\frac{1}{4} = \frac{1}{\sqrt{N_{taste}}}$ serve to divide out 
traces over taste space. The scalar current in eq.(\ref{scalarc}) is a taste 
singlet current.  Carrying out the contractions over fermionic fields in 
$C^{3pnt}$ and using the well known relation between naive quark 
propagators $G_\Psi(x,y)$ and one-component field 
propagators $G_\chi(x,y)$,
\be
G_\Psi(x,y) = \Omega(x) \Omega^\dagger(y) G_\chi(x,y)
\ee
with
\be 
\Omega(x) = \prod^3_{\mu = 0} (\gamma_\mu)^{x_\mu},
\ee
one obtains,
\begin{eqnarray}
& &\langle \Phi_K(x) \,\tilde{S}(z) \, \Phi^\dagger_D(y)
 \rangle  \nl
& = &  \frac{1}{16} Tr\left \{ G_{\Psi,s}(x,z) G_{\Psi,c}(z,y) \gamma_5 
G_{\Psi,l}(y,x) \gamma_5 \right \} \nl
 & & \nl
& = & \frac{1}{16} Tr \{ \left [ \Omega(x) \Omega^\dagger(z) 
\Omega(z) \Omega^\dagger(y) \gamma_5 \Omega(y) \Omega^\dagger(x) \gamma_5 
\right ] \nl
 & & \quad \times G_{\chi,s}(x,z) G_{\chi,c}(z,y) G_{\chi,l}(y,x)  \} \nl
& = & \frac{1}{4} \phi(y) \phi(x)
tr \left \{ G_{\chi,s}(x,z) G_{\chi,c}(z,y) G_{\chi,l}(y,x) \right \} \nl
& = & \frac{1}{4} \phi(y) \phi(z)
tr \left \{ G^\dagger_{\chi,s}(z,x) G_{\chi,c}(z,y) G_{\chi,l}(y,x) \right \}. \nl
& & 
\end{eqnarray}
$\phi(y)$ stands for $(-1)^{\sum_\mu y_\mu}$
 and similarly for $\phi(x)$  and $\phi(z)$. 
In the last step we have used $G_\chi(x,y) = \phi(x) * \phi(y) 
G^\dagger_\chi(y,x)$.  
``Tr'' is the trace over spin and color and ``tr'' the trace only over color.
The three-point correlator can now be written as
\begin{eqnarray}
\label{thrpnt2}
& & C^{3pnt}(t_0,t,T,\vec{p}_K) = 
\frac{1}{L^3} \sum_{\vec{x}}
\sum_{\vec{y}} \sum_{\vec{z}} e^{i\vec{p}_K \cdot (\vec{z} - \vec{x})}  \times \nl
&  & \frac{1}{4} \phi(y) \phi(z)
 \langle tr
 \left \{ G^\dagger_{\chi,s}(z,x) G_{\chi,c}(z,y) G_{\chi,l}(y,x) \right \}
 \rangle, \nl
\end{eqnarray}
with $x_0 \equiv t_0$, $y_0 \equiv t_0 - T$ and $z_0 \equiv t$ and where 
$\langle \rangle$ now stands for average over configurations.
  One sees from 
eq.(\ref{thrpnt2}) that strange HISQ propagators are needed going 
from $(\vec{x},t_0)$ to general $z$ and light propagators again from 
$(\vec{x},t_0)$ to general $y$.  If one actually wanted to carry out the 
$\frac{1}{L^3} \sum_{\vec{x}}$  one would need a strange and a light 
propagator from each spatial point on time slice $t_0$ and that would 
be prohibitively expensive.   A common approach is to give up on doing 
the $\sum_{\vec{x}}$ and to use ``local sources'' where $\vec{x}$ is fixed 
at some $\vec{x}_0$, e.g. $\vec{x} = \vec{0}$. One then has,
\begin{eqnarray}
\label{thrpntloc}
& & C_{loc}^{3pnt}(t_0,t,T,\vec{p}_K) = 
\sum_{\vec{y}} \sum_{\vec{z}} e^{i\vec{p}_K \cdot \vec{z}}  \times \nl
&  & \frac{1}{4} \phi(y) \phi(z)
 \langle tr \left \{ G^\dagger_{\chi,s}(z,x_{loc}) G_{\chi,c}(z,y)
 G_{\chi,l}(y,x_{loc})
 \right \} \rangle, \nl
\end{eqnarray}
with $x_{loc} = (\vec{0},t_0)$.  
Momentum conservation will ensure that only kaons with momentum $\vec{p}_K$ 
 contribute and be picked out at time slice $t_0$.
On the other hand ``random wall'' sources allow us to carry out the 
$\frac{1}{L^3} \sum_{\vec{x}}$ without having to invert at each spatial 
point.  This can be seen by writing,
\begin{eqnarray}
\label{thrpntrw}
& & C_{rw}^{3pnt}(t_0,t,T,\vec{p}_K) = 
\frac{1}{L^3} \sum_{\vec{x}} \sum_{\vec{x}^\prime} 
\sum_{\vec{y}} \sum_{\vec{z}} e^{i\vec{p}_K \cdot (\vec{z} - \vec{x})} \times  \nl
&  & \frac{1}{4} \phi(y) \phi(z)
 \langle tr \{ G^\dagger_{\chi,s}(z,x) G_{\chi,c}(z,y) G_{\chi,l}(y,x^\prime)
  \nl
& & \qquad \qquad \qquad \qquad \qquad \qquad \times \;
 \xi^*(\vec{x}) \xi(\vec{x}^\prime) \} \rangle 
\end{eqnarray}
 $\xi(\vec{x})$ is a field of  random U(1) phases, and $\langle \xi^*(\vec{x}) 
\xi(\vec{x}^\prime) \rangle = \delta_{\vec{x},\vec{x}^\prime}$ 
ensures that (\ref{thrpntrw}) reduces to (\ref{thrpnt2}) after 
averaging over gauge field configurations.  So in the random wall source 
approach one calculates the light quark propagator $G_{\chi,l}$ using the source 
$\frac{1}{\sqrt{L^3}}\sum_{\vec{x}^\prime}\xi(\vec{x}^\prime)$ 
and $G_{\chi,s}$, the strange propagator 
by using the source 
$ \frac{1}{\sqrt{L^3}}\sum_{\vec{x}}\xi(\vec{x}) e^{i \vec{p}_K \cdot \vec{x}}$. 
Only one light quark inversion is required in addition to a separate 
strange quark inversion for each $\vec{p}_K$. In this way one obtains 
the random wall propagators,
\be
\label{rwq}
G_{\chi,l}^{rw}(y,t_0) \equiv \frac{1}{\sqrt{L^3}} \sum_{\vec{x}^\prime} 
G_{\chi,l}(y,x^\prime) \xi(\vec{x}^\prime)
\ee
and
\be
\label{rwstrange}
G_{\chi,s}^{rw}(z,t_0;\vec{p}_K) \equiv \frac{1}{\sqrt{L^3}} \sum_{\vec{x}} 
G_{\chi,s}(z,x) \xi(\vec{x}) e^{i \vec{p}_K \cdot \vec{x}}.
\ee
The expression for the three-point correlator becomes,
\begin{eqnarray}
\label{thrpnt3}
& & C_{rw}^{3pnt}(t_0,t,T,\vec{p}_K) = 
\sum_{\vec{y}} \sum_{\vec{z}} e^{i\vec{p}_K \cdot \vec{z}} \times \nl
&  & \frac{1}{4} \phi(y) \phi(z)
 \langle tr \left \{ G^{rw \dagger}_{\chi,s}(z,t_0;\vec{p}_K) G_{\chi,c}(z,y)
 G^{rw}_{\chi,l}(y,t_0)
 \right \} \rangle, \nl
\end{eqnarray}
The charm propagator in eq.(\ref{thrpnt3}) is obtained by inverting at 
time slice $y_0 = t_0 - T$ with source $\sum_{\vec{y}} \phi(y) G^{rw}_{\chi,l}(y,t_0)$.  
In this way one gets the ``sequential'' charm propagator,
\be
\label{seqcharm}
G_{\chi,c}^{seq}(z,t_0,T) \equiv  \sum_{\vec{y}} \phi(y) 
G_{\chi,c}(z,y) G_{\chi,l}^{rw}(y,t_0),
\ee
and
\begin{eqnarray}
\label{thrpnt4}
& & C_{rw}^{3pnt}(t_0,t,T,\vec{p}_K) =  
 \sum_{\vec{z}} e^{i\vec{p}_K \cdot \vec{z}} \times \nl
& & \frac{1}{4}  \phi(z)
 \langle tr \left \{ G^{rw \dagger}_{\chi,s}(z,t_0;\vec{p}_K) G^{seq}_{\chi,c}
(z,t_0,T) \right \} \rangle.
\end{eqnarray}
The most costly part of our simulations is calculating the random wall strange quark 
propagators of eq.(\ref{rwstrange}). A separate inversion is required 
for each $\vec{p}_K$ (e.g. 8 inversions for the different 
combinations $(\pm 1,\pm 1,\pm 1)$). On the other hand, when we change the $T$ values 
only $G^{seq}_{\chi,c}$ of eq.(\ref{seqcharm}) needs to be recalculated 
and one inversion suffices for all momenta.  This is one of the reasons why the full 
kaon momentum $\vec{p}_K$ is put into (\ref{rwstrange}) and none into 
(\ref{rwq}).

\begin{figure}
\includegraphics*[width=8cm,height=9cm,angle=-90]{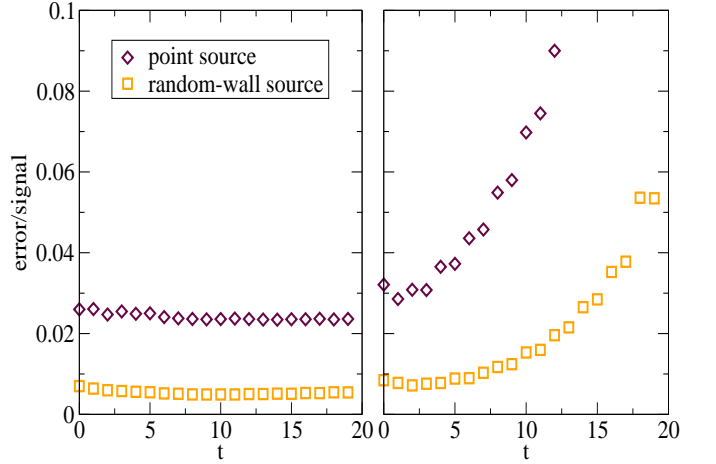}
\caption{
Comparison between local and random wall sources for $\vec{p} = \frac{2 \pi}{La} 
(0,0,0)$ (left plot), and for $\vec{p} = \frac{2 \pi}{La} 
(1,1,1)$ (right plot).
 }
\label{ranE}
\end{figure}

\vspace{.05in}
In Fig.~\ref{ranE} we show 
comparisons of percentage errors in three-point correlator data 
of local sources versus random wall sources.  One sees significant 
improvement coming from random wall sources. These tests were carried out in the 
test case $D_s \rightarrow \eta_s, l \nu$ calculations and with less than 
the full statistics. 
  For $D \rightarrow K, 
l \nu$ we immediately went to random wall sources.

\vspace{.05in}
In addition to the three-point correlators, as we will show in the next section, 
several two-point correlators are needed in order to extract the matrix element 
$\langle K | S | D\rangle$.  They are,
\be
C^{2pnt}_D(t,t_0) =\frac{1}{L^3} \sum_{\vec{x}} \sum_{\vec{y}} 
\langle \Phi_D(\vec{y},t) \Phi^\dagger_D(\vec{x},t_0) \rangle,
\ee
and
\begin{eqnarray}
 & &C^{2pnt}_K(t,t_0; \vec{p}_K) = \nl 
& & \quad \frac{1}{L^3} \sum_{\vec{x}} \sum_{\vec{y}} 
e^{i \vec{p}_K \cdot (\vec{x} - \vec{y})}\langle \Phi_K(\vec{y},t)
 \Phi^\dagger_K(\vec{x},t_0) \rangle.
\end{eqnarray}
The $\frac{1}{L^3} \sum_{\vec{x}}$ can again be implemented via random wall 
sources.  As already mentioned in the previous section we also calculated 
correlators for the $\eta_c$, $\eta_s$ and $D_s$ mesons in exactly the same way as 
$C^{2pnt}_{D/K}$ in order to carry out and check mass tunings.
We have accumulated simulation data for the three- and two-point correlators 
described in this section for the ensembles of Table~\ref{T.lat}.  In the next section 
we explain how hadronic matrix elements such as $\langle K | S | D \rangle$ and 
meson masses and decay constants are extracted from this data.

\section{Fits and Data Analysis}
The interpolating operators $\Phi_D$ and $\Phi_K$ do not create just the ground state 
$D$ meson or kaon that we are interested in but also excited states with the 
same quantum numbers.  With staggered quarks there is the further
complication that in addition to regular states so-called ``parity partner'' 
states can contribute, whose energies are measured relative 
to $i \pi$ so that $e^{- E t} \rightarrow (-1)^t e^{ - Et}$.  The $D$ meson 
correlator, for instance, has the $t$ dependence (we set $t_0 = 0$ for simplicity),
\begin{eqnarray}
\label{twopntfit}
C^{2pnt}_D(t) &=& \sum_{j=0}^{N_D-1} b^D_j (e^{-E_j^D t} + e^{-E_j^D ( N_t - t)}) \nl
&+& \sum_{k=0}^{N_D^{\prime} - 1} d^D_k (-1)^t 
( e^{-E^{\prime D}_k t} + e^{-E_k^{\prime D} (N_t - t)}).\nl
\end{eqnarray}
We are interested in the ground state $D$ meson contribution with amplitude,
\be
\label{b0d}
b^D_0 \equiv \frac{|\langle \Phi_D | D \rangle|^2}{2 M_D a^3}.
\ee
Similar relations apply for other mesons.  Only in the case 
of equal mass mesons ($\pi$, $\eta_s$ or $\eta_c$) at zero momentum are 
the oscillatory contributions absent.
The three-point correlators such as eq.(\ref{thrpnt4}) will have contributions 
from regular and oscillatory states for both the kaon and the $D$ meson.  
The rather complicated $t$ and $T$ dependence is then given by,
\begin{eqnarray}
\label{thrpntfit}
& &C^{3pnt} (t,T)
 = \sum_j^{N_K-1} \sum_k^{N_D-1} A_{jk} e^{-E_j^K t} e^{ -E_k^D (T-t)} \nl
& & \quad + \sum_j^{N_K-1} \sum_k^{N_D^\prime - 1} B_{jk}
 e^{-E_j^K t} e^{ -E_k^{\prime D} (T-t)} (-1)^{(T-t)} \nl
& & \quad + \sum_j^{N_K^\prime - 1} \sum_k^{N_D-1} C_{jk}
 e^{-E_j^{\prime K} t} e^{ -E_k^D (T-t)} (-1)^t \nl
& & \quad + \sum_j^{N_K^\prime - 1} \sum_k^{N_D^\prime - 1} D_{jk} e^{-E_j^{\prime K} t}
 e^{ -E_k^{\prime D} (T-t)} (-1)^t (-1)^{(T-t)}. \nl
\end{eqnarray}
We will only consider the region $0 \leq t \leq T$ and take $T << N_t$ so that 
any contributions from mesons propagating ``around the lattice'' due to periodic 
boundary conditions in time can be neglected.
The relevant amplitude here is,
\be
\label{a00}
A_{00} \equiv \frac{\langle \Phi_K|K\rangle \, \langle K|S|D\rangle \, \langle D|\Phi_D \rangle}
{(2 E_K a^3) \, (2 M_D a^3)} \, a^3.
\ee
From eqs.(\ref{b0d}) and (\ref{a00}) one sees that our sought after hadronic matrix 
element is given by,
\be
\label{ksd}
\langle K | S | D \rangle = 2 \sqrt{ M_D E_K} \; \frac{A_{00}}{\sqrt{b_0^K b_0^D}}.
\ee
So our goal is to extract the combination on the right-hand-side of (\ref{ksd}) 
as accurately as possible and with any correlations among the errors of the 
individual components, $A_{00}$, $b_0^{K/D}$, $M_D$ and $E_K$ taken properly 
into account.  

\vspace{.05in}
We have carried out simultaneous fits to $C^{2pnt}_D$, $C^{2pnt}_K$ and 
the three-point correlators with different $T$ values, 
$C^{3pnt}(t,T_i)$ $i = 1,2 ...$. following the fit ansaetze of eqs.(\ref{twopntfit}) 
and (\ref{thrpntfit}). 
Two (three) different $T$ values are used for the coarse (fine) ensembles.
This allows us to evaluate (\ref{ksd}) within one fit.
The two-point correlators were fit for $t$-values between $t_{min} = 2 (2)$ and 
$t_{max} = 30 (20 \sim 30)$ for the $K$ and $D$ respectively for coarse lattices.
For fine lattices, $t$-values were used between $t_{min}= 2 \sim 4 (2)$ and 
$t_{max} = 30 (30)$.
  For the three-point correlators 
we used all the data between $t = 2$ and $t = T - 2$ for coarse lattices, and $t = 3$ and $t = T - 2$ for fine lattices. 
Simultaneous fits with multiple $T$ and taking different fit ranges for different correlators were also helpful to reduce the statistical errors, since it allows us to extract maximum information from both the three-point and two-point correlators.
The number of exponentials 
in our fit ansatz was varied to test for stability of fit results. For our final fits
we ended up choosing around $3 \sim 4$ for $N_D$ and $N_K$.  $N_{D/K}^\prime$ was taken 
to be mostly $N_{D/K} - 1$.
  Figs \ref{a00} - \ref{bK} show some results for 
$A_{00}$, $M_D$, $b^D_0$, $E_K$, $b^K_0$ versus $N_D$ or $N_K$ for ensemble C1 at kaon momentum $\vec{p}=(0,0,0)$.  
All our fits are carried out using Bayesian methods~\cite{bayes}. We describe choices for priors and 
prior widths in the Appendix. 

\begin{figure}
\includegraphics*[width=7.0cm,height=8.0cm,angle=-90]{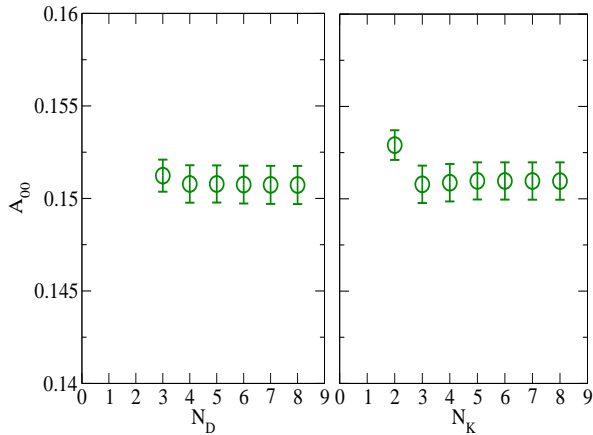}
\caption{
$A_{00}$ versus the number $N_{D/K}$. In the left (right) plot $N_K$ ($N_D$) is fixed at 3 (4).
 }
\label{a00}
\end{figure}
\begin{figure}
\includegraphics*[width=7.0cm,height=8.0cm,angle=-90]{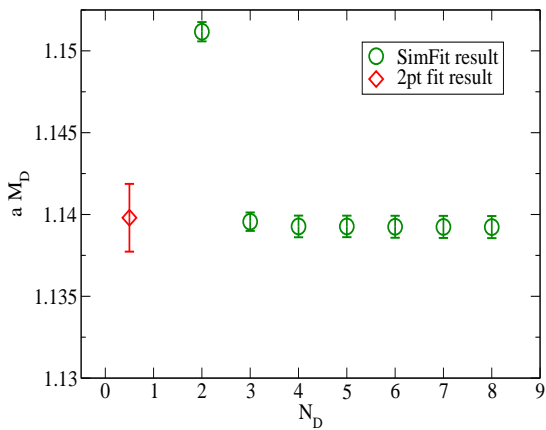}
\caption{
$aM_D$ versus $N_D$. Green circles are from simultaneous $C^{2pnt} - C^{3pnt}$ fits. The red diamond is from fits to just the $C^{2pnt}$.
 }
\label{mD}
\end{figure}
\begin{figure}
\includegraphics*[width=7.0cm,height=8.0cm,angle=-90]{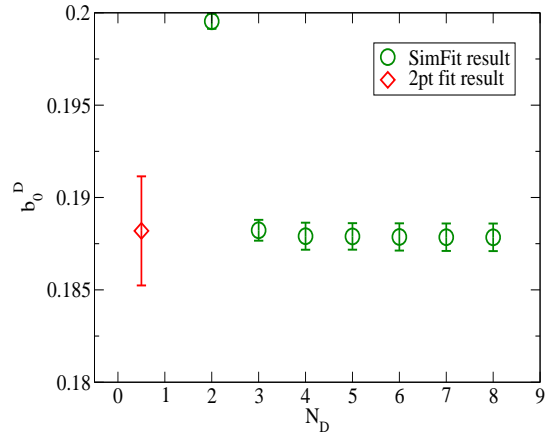}
\caption{
Same as Fig.~\ref{mD} for $b_0^D$.
 }
\label{bD}
\end{figure}
\begin{figure}
\includegraphics*[width=7.0cm,height=8.0cm,angle=-90]{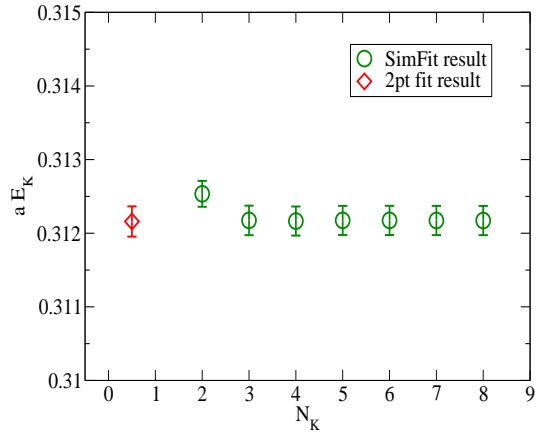}
\caption{
Same as Fig.~\ref{mD} for $aE_K$ versus $N_K$.
 }
\label{mK}
\end{figure}
\begin{figure}
\includegraphics*[width=7.0cm,height=8.0cm,angle=-90]{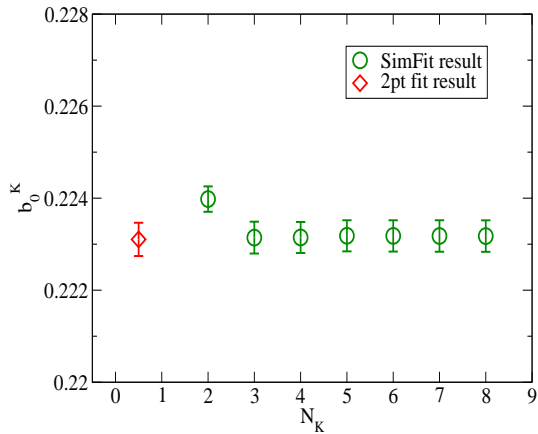}
\caption{
Same as Fig.~\ref{mD} for $b_0^K$ versus $N_K$.
 }
\label{bK}
\end{figure}

\begin{figure}
\includegraphics*[width=7.0cm,height=8.0cm,angle=-90]{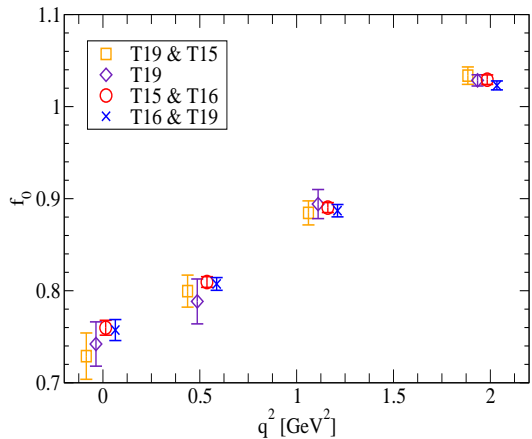}
\caption{
Effect of multi T-fits
 }
\label{mulT}
\end{figure}

\vspace{.05in}
  We have found that 
using data from several $C^{3pnt}$ with different $T$-values helps greatly 
in reducing statistical/fitting errors. Fig.~\ref{mulT} compares results for $f_0(q^2)$ for 
ensemble C2.  One sees that having two rather than just one $C^{3pnt}(t,T_i)$ 
involved in the simultaneous fit reduces errors and that this effect is most 
pronounced when one combines an even $T$ with an odd $T$.  It may not be surprising 
that improvements are achieved from multi-T fits. 
 From the fit ansatz (\ref{thrpntfit}) one 
sees that having more $C^{3pnt}$'s does not increase the number of 
fit parameters ($A_{jk}$ etc.) although the amount of data and hence of 
information given to the minimizer is increased.  Furthermore if $T_1 + T_2$ 
is odd, then $(-1)^{T_1-t}$ and $(-1)^{T_2 -1}$  have opposite signs and 
$C^{3pnt}(t,T_1)$ and $C^{3pnt}(t,T_2)$ will provide more independent 
information. 

\begin{figure}
\includegraphics*[width=7.0cm,height=8.0cm,angle=-90]{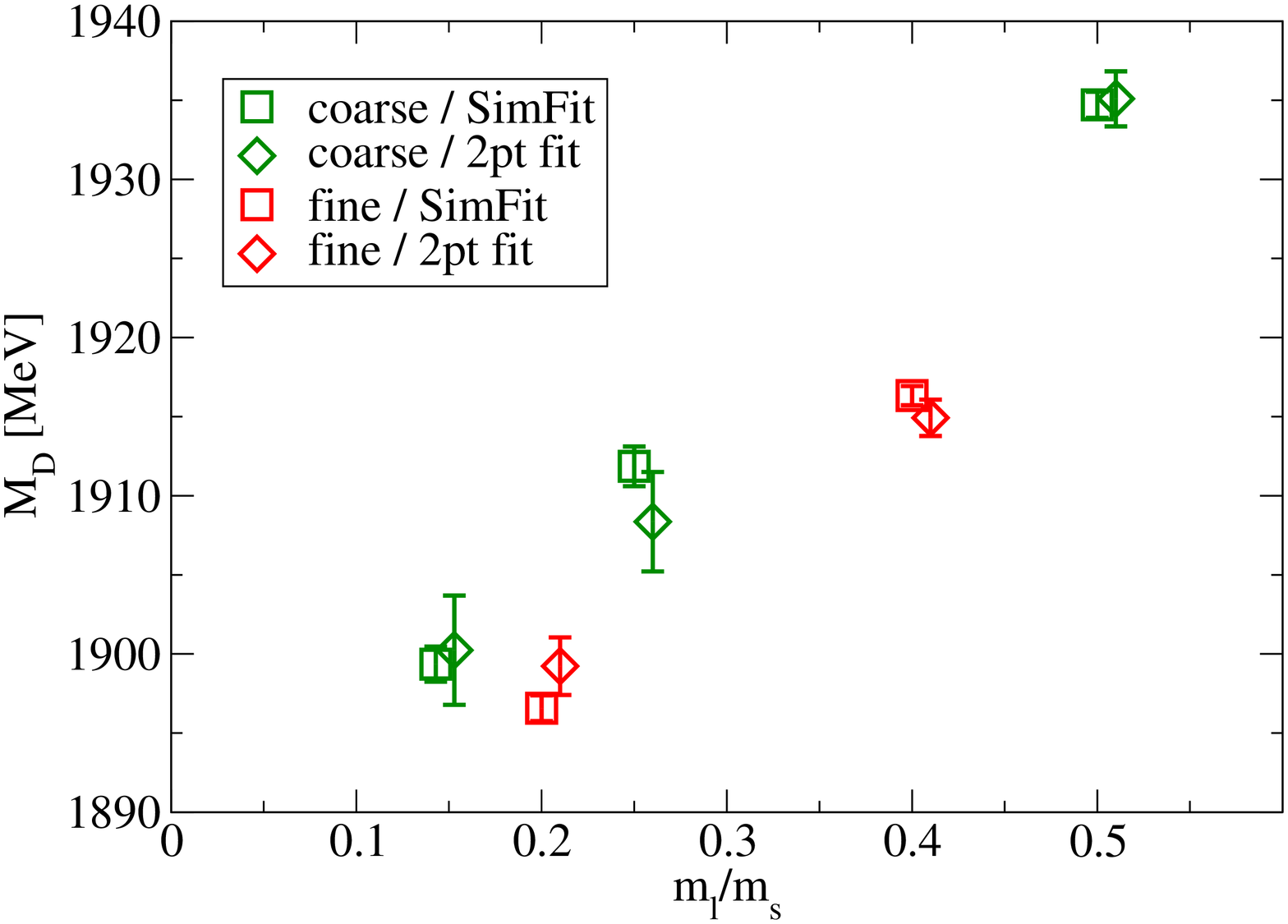}
\caption{
Comparison of $M_D$ from two-point and simultaneous fits
 }
\label{mD2}
\end{figure}
\begin{figure}
\includegraphics*[width=7.0cm,height=8.0cm,angle=-90]{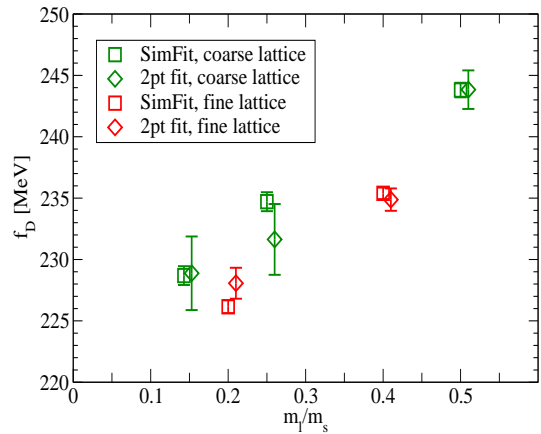}
\caption{
Comparison of $f_D$ from two-point and simultaneous fits
 }
\label{fD}
\end{figure}

\vspace{.05in}
Most of the fit parameters such as $M_D$ or $b_0^D$ etc. that one gets from 
the simultaneous $C^{2pnt}$ - $C^{3pnt}$ fits can also be determined by 
just fitting the two-point correlators by themselves.  Results from the 
$C^{2pnt}$ fits are also shown on Figs.~\ref{mD} - \ref{bK} and provide  consistency 
checks.  
One interesting outcome is that two-point correlator parameters are 
more accurately determined via simultaneous fits with three-point 
correlators than when they are fit alone.  This is especially 
noticeable for $D$ meson correlators, namely for $M_D$ and $b_0^D$ and 
has implications for determinations of the decay constant $f_D$. The latter is
related to $b_0^D$ through,
\be
\label{fd}
a f_D = \frac{m_{0,c} + m_{0,l}}{M_D} \sqrt{\frac{2 b_0^D}{ a M_D}}.
\ee
In Figs.~\ref{mD2} and \ref{fD} we compare results for $M_D$ and $f_D$ using 
either pure two-point fits or simultaneous fits. One sees the significant 
improvement coming from the simultaneous fit. 
Doing simultaneous fits gives a better handle on excited state contributions
because they contribute differently to two-point and three-point correlators.
This is effectively similar to adding smearings to the correlators.
In section VIII we will discuss extracting $f_D$ in the chiral/continuum limit.  
Having  simultaneous fit results will make this determination more accurate 
than what can be achieved from pure two-point correlators.  This appears to 
be a bonus side product of semileptonic decay studies.  

\begin{figure}
\includegraphics*[width=7.0cm,height=8.0cm,angle=-90]{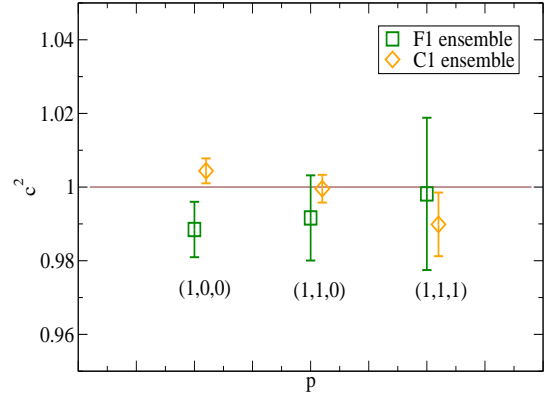}
\caption{
Speed of light squared versus momentum for the kaon from the 2-pt fit
 }
\label{sol}
\end{figure}

\vspace{.05in}
In Tables~\ref{T.mdfd}, \ref{T.mkfk} and \ref{T.f0} we summarize our main fit results.  One sees from Table~\ref{T.f0}
that we were able to determine $f_0(\vec{p}_K)$ with errors ranging from 
$\sim 0.2$\% at zero momentum to $\sim 0.9$\% at our highest momentum.
In Fig.~\ref{sol}
we plot the square of the ``speed of light'' $c^2(\vec{p})$ for the kaon.  
\be
c^2(\vec{p}) = \frac{E^2_K(\vec{p}) - M^2_K}{\vec{p}^2}
\ee
One sees that the relativistic dispersion relation is satisfied 
to about $1 \sim 2$\%.  We will check the effect of deviations from 
exact continuum dispersion relations 
on our final results for form factors in later sections.

\begin{table}
\caption{
Fit results for $aM_D$, $af_D$, $aM_{D_s}$ and $af_{D_s}$.  Numbers 
in brackets are from two-point fits, whereas the rest come from 
simultaneous fits.
}
\label{T.mdfd}
\begin{center}
\begin{tabular}{|c|c|c|c|c|}
\hline
Set & $aM_D$   & $af_D$ &  $aM_{D_s}$ & $af_{D_s}$ \\
\hline
\hline
C1  & 1.1393(7)&0.1372(4) & &  \\
    & (1.1398(21)) & (0.1373(18))  & (1.1876(5)) & (0.1539(6)) \\
C2  &1.1595(8)&0.1423(4)& &  \\
    & (1.1574(19)) & (0.1405(17))  & (1.2008(6)) & (0.1560(7)) \\
C3  &1.1618(5)&0.1464(3)& &  \\
    & (1.1620(10)) & (0.1464(9))  & (1.1899(5)) & (0.1553(4)) \\
\hline
F1  &0.8141(3)&0.0971(2)& &  \\
    & (0.8152(8)) & (0.0979(5))  & (0.8473(2)) & (0.1083(2)) \\
F2  & 0.8197(3)&0.1007(2)& &  \\
    & (0.8191(5)) & (0.1005(4))  & (0.8435(2)) & (0.1078(2)) \\
\hline
\end{tabular}
\end{center}
\end{table}

\begin{table}
\caption{
Fit results for $aM_K$, $aE_K(\vec{p})$, $af_K$, $am_\pi$ and $af_\pi$.  Numbers 
in brackets are from two-point fits, whereas the rest come from 
simultaneous fits.
}
\label{T.mkfk}
\begin{center}
\begin{tabular}{|c|c|c|c|c|}
\hline
Set & $aM_K$ & $aE_K$ & $aE_K$& $aE_K$  \\
    &   &  (1,0,0)   & (1,1,0) &  (1,1,1)  \\
\hline
\hline
C1  &0.3122(2) &0.4081(7) &0.4837(9) &0.5469(20)   \\
    & (0.3122(2)) & (0.4083(7))  & (0.4842(9)) & (0.5477(24))  \\
C2  &0.3285(5) &0.4531(16) &0.5525(17) &0.6373(32)   \\
    & (0.3285(3)) & (0.4536(13))  & (0.5532(18)) & (0.6382(33)) \\
C3  &0.3572(2) &0.4750(9) &0.5720(10) &0.6524(22)   \\
    & (0.3572(2)) & (0.4755(9))  & (0.5722(11)) & (0.6524(35)) \\
\hline
F1  &0.2285(2) &0.3203(7) &0.3919(9) &0.4559(15)   \\
    & (0.2286(2)) & (0.3185(12))  & (0.3896(23)) & (0.4506(47)) \\
F2  &0.2460(1) &0.3340(4) &0.4014(7) &0.4609(11)   \\
    & (0.2458(2)) & (0.3334(7))  & (0.4015(10)) & (0.4616(16)) \\
\hline
\end{tabular}
\begin{tabular}{|c|c|c|c|}
\hline
Set &  $af_K$ & $am_\pi$ & $af_\pi$ \\
\hline
\hline
C1   &0.1011(1) & & \\
     & (0.1011(1))& (0.1599(2))    & (0.0893(1)) \\
C2   &0.1044(1) & & \\
     & (0.1045(1))& (0.2108(2))     & (0.0949(1)) \\
C3   &0.1079(1) & & \\
     & (0.1079(1)) & (0.2931(2))      & (0.1023(1)) \\
\hline
F1 &0.0721(1) & & \\
     & (0.0721(1))& (0.1344(2))      & (0.0645(1))\\
F2 &0.0748(1) & & \\
     & (0.0747(1)) & (0.1873(1))      & (0.0697(1))\\
\hline
\end{tabular}
\end{center}
\end{table}

\begin{table}
\caption{
Fit results for $f_0(\vec{p}_K)$ 
}
\label{T.f0}
\begin{center}
\begin{tabular}{|c|c|c|c|c|}
\hline
Set & $f_0(0,0,0)$ & $f_0(1,0,0)$ & $f_0(1,1,0)$& $f_0(1,1,1)$  \\
\hline
\hline
C1  &1.022(3) &0.916(3) &0.846(3) &0.794(6)  \\
C2  &1.023(4) &0.885(5) &0.807(4) &0.758(7)  \\
C3  &1.010(2) &0.883(3) &0.803(4) &0.754(5)  \\
\hline
F1  &1.019(2) &0.876(3) &0.796(3) &0.745(4)  \\
F2  &1.011(1) &0.874(2) &0.792(2) &0.739(4)  \\
\hline
\end{tabular}
\end{center}
\end{table}

\section{Chiral and Continuum Extrapolations Using the $z$-Expansion}
The twenty entries in Table~\ref{T.f0} summarize our results for the form factor 
$f_0(q^2)$ evaluated on the five ensembles of Table~\ref{T.lat} with 
four different momenta $\vec{p}_K$ (including zero momentum) 
per ensemble. The kaon energy in 
the $D$ rest frame, $E_K$, is related to $q^2$ via,
\be
q^2 = M_D^2 + M_K^2 - 2 M_D E_K,
\ee
and the physical region is $0 \leq q^2 \leq q^2_{max} = (M_D - M_K)^2$. 
 $M_K$, $E_K$ and $M_D$ are given  for the different ensembles in 
Tables~\ref{T.mdfd} \& \ref{T.mkfk}.
The next step is to extrapolate the data of Table~\ref{T.f0} to the chiral/continuum 
physical limit. 
As is well known, chiral extrapolations for form factors are much more subtle 
than for static quantities such as masses or decay constants. The main reason for 
this is that form factors depend not only on meson/quark masses but also 
on a kinematic variable such as $q^2$ (or equivalently on $E_K$).  Kinematic 
variables are themselves functions of meson masses.  It is not sufficient to 
parameterize just the light quark mass dependence of form factors. 
  One must at the same time capture the kinematic variable 
dependence correctly for each value of the light quark mass and the lattice 
spacing (i.e. for each of our ensembles). 
  Furthermore we are interested in a parameterization that works over
 the entire physical kinematic range. In the chiral limit, $q^2$ ranges between 
$0 \leq q^2 \leq 1.9 {\rm GeV}^2$ and $E_K$ between $0.495 \leq E_K \leq 1.0$GeV. 

\vspace{.1in}
\noindent
\underline{ The $z$-expansion}

\vspace{.05in}
In addition to $q^2$ and $E_K$ a third kinematic variable has proven to be convenient 
 in semileptonic form factor studies, in particular in recent analysis of 
$B \rightarrow \pi, l \nu$ decays \cite{boyd, arnesen, hill}.
\be
\label{zparam}
z(q^2,t_0) = \frac{\sqrt{t_+ - q^2} - \sqrt{t_+ - t_0}}
{\sqrt{t_+ - q^2} + \sqrt{t_+ - t_0}}, 
\ee
where $t_\pm = (M_D \pm M_K)^2$ and $t_0$ is a free parameter 
that defines the zero of the $z$-variable, $z(q^2 = t_0,t_0) = 0$.
By going to the $z$-variable one is mapping the cut region $t_+ < q^2 < \infty$ 
in the complex $q^2$ plane onto the circle $|z| = 1$ and $ -\infty < q^2 < t_+$ 
onto $z \in [-1,1]$.  The physical region $0 \leq q^2 \leq q^2_{max} = t_-$ 
corresponds then to an even smaller region around $z=0$.  For instance for 
the choice $t_0 = 0.5 t_-$ and for physical values of $M_D$ and $M_K$ 
one has $-0.057 \leq z \leq 0.046$. In other words one always has $|z| < 0.06$ 
in the physical region and this should make $z$ a good variable for a 
power series expansion.  As discussed in the literature using 
analyticity properties of form factors one can write,
\be
\label{f0zexp}
f_0(q^2) = \frac{1}{P(q^2) \, \Phi_0(q^2,t_0)} \sum_{k = 0}^\infty 
a_k(t_0) z(q^2,t_0)^k.
\ee
The function $P(q^2)$ in the denominator is there to factor out any 
isolated poles in the region $t_- < q^2 < t_+$ below the 
$DK$ threshold at $q^2 = t_+$.  In the case of $D \rightarrow K$ 
semileptonic decays the charm-strange scalar current has the same quantum numbers  
as the $D^*_{s0}(2317)$ $0^+$ meson so that one choice for $P(q^2)$ would be 
$P(q^2) = (1 - q^2/(M_{D_{s0}^*})^2)$.  We have worked with both $P(q^2) = 
(1 - q^2/(M_{D_{s0}^*})^2)$ and $P(q^2) = 1$ and find that 
although the expansion coefficients $a_k$  depend on the choice for $P(q^2)$, 
 in either case the data can be reproduced very well with 
just a few terms (as we will discuss below, with just three terms) 
in the $z$-expansion.

\vspace{.05in}
For the ``outer function'' $\Phi_0$ we adopt the choice given in 
Ref. \cite{arnesen}.
\begin{eqnarray}
\Phi_0(q^2,t_0) &=&  \sqrt{\frac{3 t_+t_-}{32 \pi \chi_0}} 
(\sqrt{t_+ - q^2} + \sqrt{t_+ - t_0}) \times \nl
& & \frac{(t_+ - q^2)^{1/2}}{(t_+ - t_0)^{1/4}} 
\frac{(\sqrt{t_+ - q^2} + \sqrt{t_+ - t_-} )^{1/2}}
{(\sqrt{t_+ - q^2} + \sqrt{t_+})^4}. \nl
\end{eqnarray}
$\chi_0$ has been calculated in the literature using QCD perturbation theory and 
the Operator Product Expansion (OPE). An expression including ${\cal O}(\alpha_s)$ 
and condensate contributions is given for instance in Ref. \cite{arnesen}. 
 For fixed charm quark mass 
$\chi_0$ is a constant and affects just the overall normalization of the 
$a_k$'s.  For simplicity we ignore the condensate contributions, which are of ${\cal O}(m_c^{-3})$ 
and ${\cal O}(m_c^{-4})$ respectively, and retain just the tree-level and ${\cal O}(\alpha_s)$
contributions.   Any other choice would just mean a common overall rescaling of the 
expansion coefficients $a_k$. 

\vspace{.1in}
\noindent
\underline{ Testing the $z$-expansion with individual fits}

\begin{figure}
\includegraphics*[width=7.0cm,height=8.0cm,angle=-90]{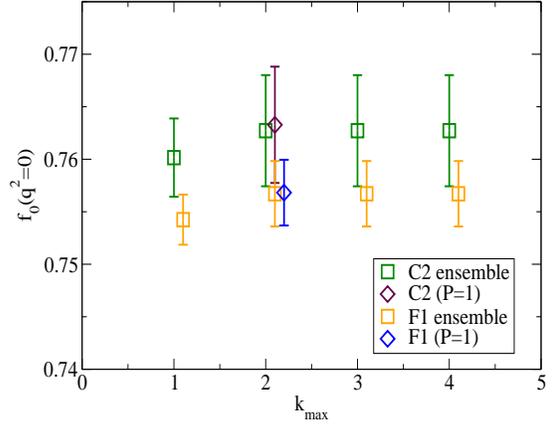}
\caption{
$f_0(0)$ versus $k_{max}$ for ensembles C2 and F1.  
$P(q^2) = (1 - q^2/(M_{D_{s0}^*})^2)$ everywhere except 
for at $k_{max} = 2$ where results for both $P(q^2) = 1$ and 
$P(q^2) = (1 - q^2/(M_{D_{s0}^*})^2)$ are shown.  
 }
\label{f0_kmax}
\end{figure}

\vspace{.05in}
In order to test the usefulness of the $z$-expansion we first fit $f_0(q^2)$ 
separately for each individual ensemble using the ansatz of eq.(\ref{f0zexp}) 
with $\sum_k \rightarrow \sum_{k = 0}^{k_{max}}$.  
In each case values used for $M_D$, $M_K$ and hence also for $t_\pm$ were those 
specific to that ensemble.  We employed a common choice for $t_0$, 
$t_0 = 0.942{\rm GeV}^2$ corresponding to $ t_0 = 0.5 \times t_-^{{\rm continuum}}$.  
As is well known from $z$-expansions in general, other choices for $t_0$ made 
no difference in resulting fit curves.  
We find that good individual fits are possible once $k_{max}$ reaches $k_{max} = 2$ 
and that fit curves are then very stable with respect to further increases in 
$k_{max}$.  In Fig.~\ref{f0_kmax} we show representative results for two ensembles C2 and F1 
for $f_0(q^2=0)$ versus $k_{max}$.  We also show results for two different 
choices, $P(q^2) = 1$ and $P(q^2) = (1-q^2/(M^2_{D_{s0}^*})^2)$.  One sees that fit results 
are very insensitive to these changes in $k_{max}$ or $P(q^2)$.   At finite lattice spacing 
we have used $M_{D_{s0}^*} = M_{D_s}^{lattice} + \delta M$ with $\delta M \equiv
[M_{D_{s0}^*}(0^+) - M_{D_s}(0^-)]_{exper.}$. 
These individual $z$-expansion fit tests demonstrate (as advocated in the literature) 
  the efficiency of $z$-expansions 
in capturing the kinematics of form factors with just a small number of parameters 
and in a model independent way.

\begin{figure}
\includegraphics*[width=7.0cm,height=8.0cm,angle=-90]{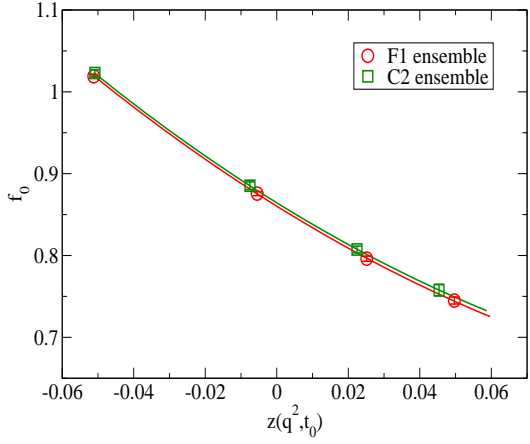}
\caption{
Individual $z$-expansion fits for ensembles C2 and F1.
 }
\label{C2_F1}
\end{figure}

In Fig.~\ref{C2_F1} we give examples of $z$-expansion fits to individual ensembles, specifically for ensembles C2 and F1.

\vspace{.05in}
\noindent
\underline{ Simultaneous modified $z$-expansion fit}

\begin{table}
\caption{
The expansion coefficients $a_i * D_i$, i = 0,1,2 from a simultaneous 
$z$-expansion fit to all data.
}
\label{T.aD}
\begin{center}
\begin{tabular}{|c|c|c|c|}
\hline
Set & $a_0 * D_0$  &  $a_1 * D_1$  & $a_2 * D_2$ \\
\hline
\hline
C1   &0.095(2) &0.085(21) &-0.07(11)  \\
C2   &0.094(2) &0.083(21) &-0.07(11)  \\
C3   &0.089(2) &0.079(21) &-0.07(11)  \\
\hline
F1   &0.094(1) &0.084(18) &-0.07(11)  \\
F2   &0.091(1)  &0.080(17) &-0.07(11)  \\
\hline
physical  &0.097(2) &0.088(18) &-0.07(11)  \\
limit     &&& \\
\hline
\end{tabular}
\end{center}
\end{table}

\begin{figure}
\includegraphics*[width=7.0cm,height=8.0cm,angle=-90]{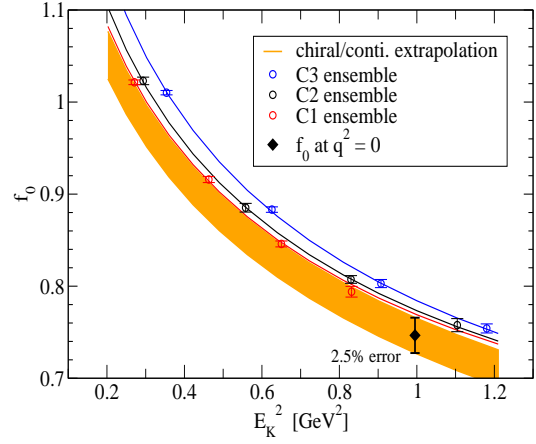}
\caption{ 
Chiral/continuum extrapolation of $f_0(q^2)$ versus $E_K^2$ from  
the modified $z$-expansion ansatz.  The data points are coarse lattice points.  
Three individual curves and the extrapolated band are from a fit to all five ensembles.  
 }
\label{coarse1}
\end{figure}

\begin{figure}
\includegraphics*[width=7.0cm,height=8.0cm,angle=-90]{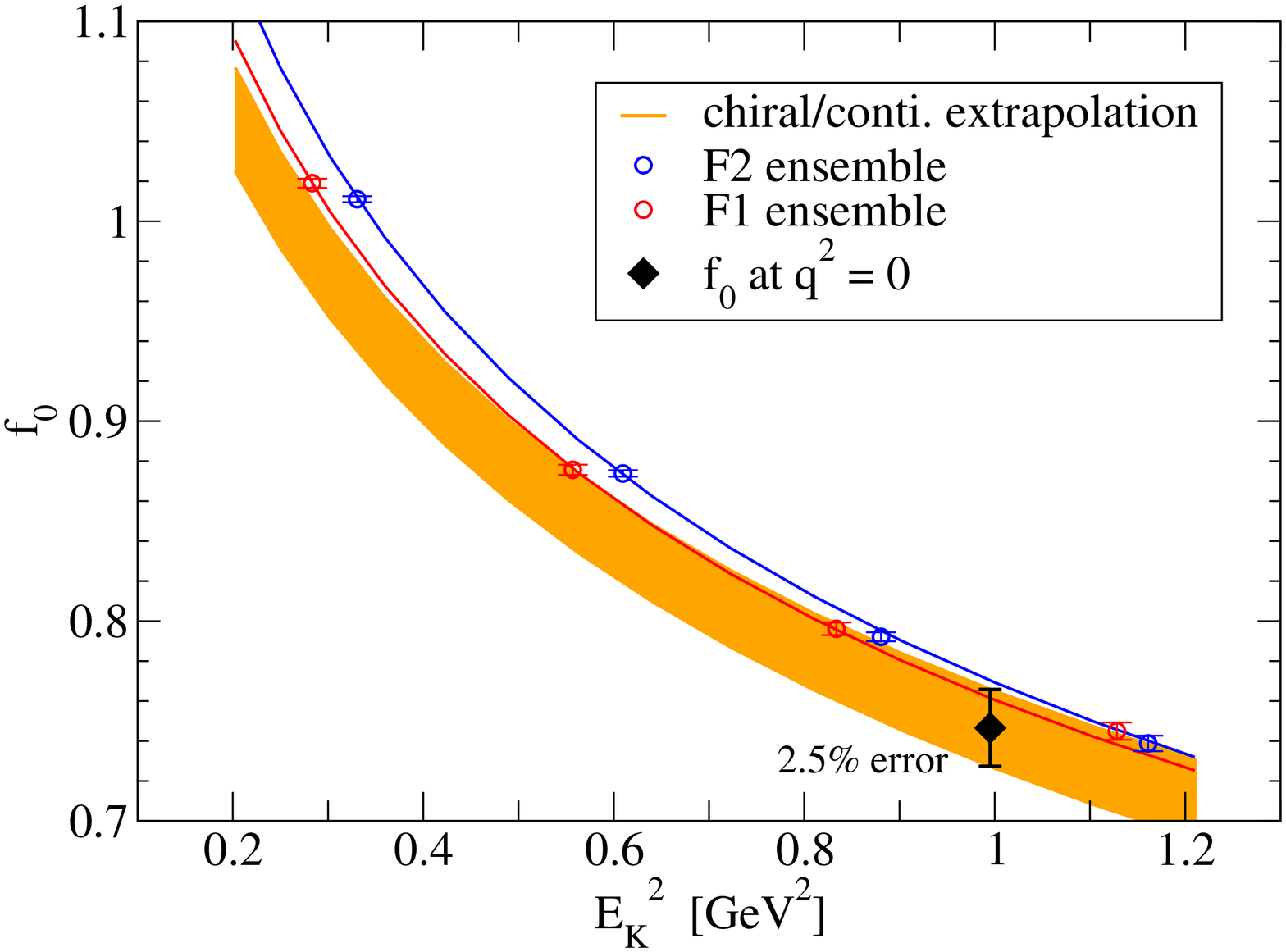}
\caption{ 
Chiral/continuum extrapolation of $f_0(q^2)$ versus $E_K^2$ from 
 the modified $z$-expansion ansatz.  The data points are fine lattice points. 
Two individual curves and the extrapolated band are from a fit to all five ensembles.
 }
\label{fine1}
\end{figure}

\begin{figure}
\includegraphics*[width=7.0cm,height=8.0cm,angle=-90]{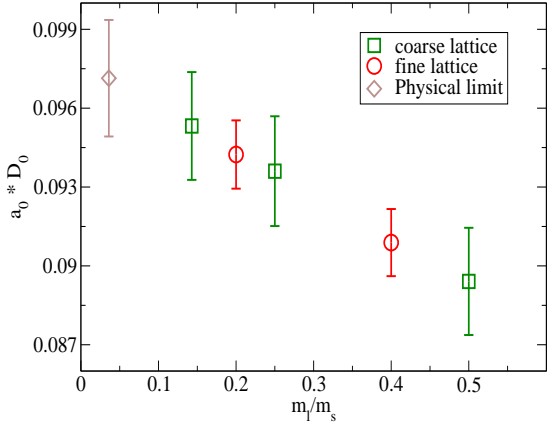}
\caption{ 
The expansion parameter $a_0 * D_0$ versus the light quark mass from a simultaneous fit to 
all data. 
 }
\label{a0}
\end{figure}

\begin{figure}
\includegraphics*[width=7.0cm,height=8.0cm,angle=-90]{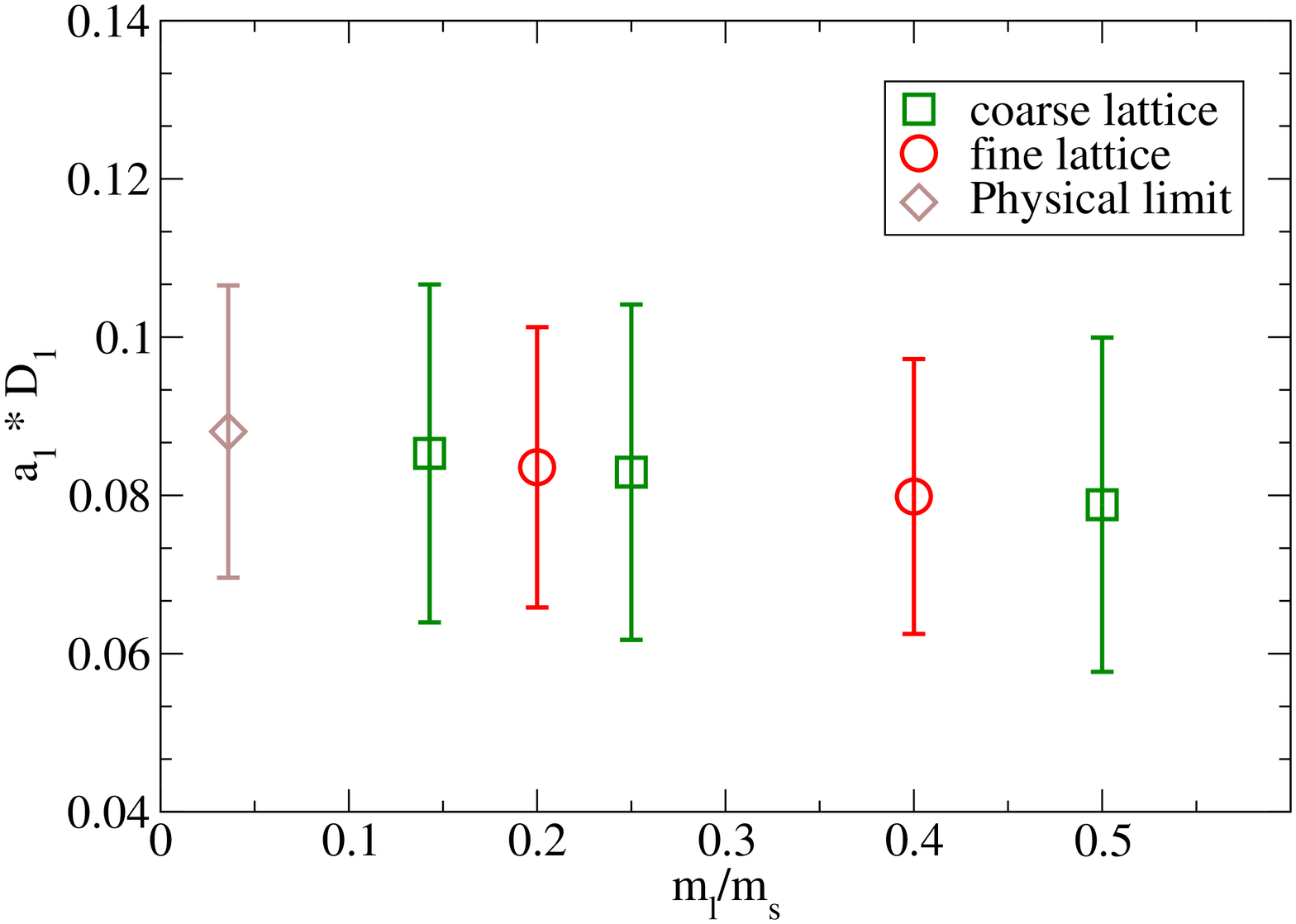}
\caption{ 
Same as Fig.~\ref{a0} for $a_1 * D_1$.
 }
\label{a1}
\end{figure}

\begin{figure}
\includegraphics*[width=7.0cm,height=8.0cm,angle=-90]{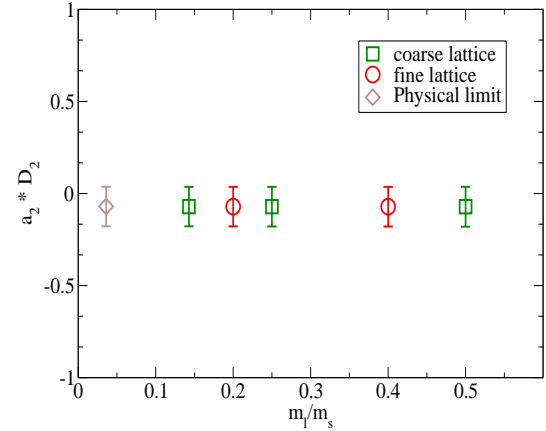}
\caption{ 
Same as Fig.~\ref{a0} for $a_2 * D_2$.
 }
\label{a2}
\end{figure}

\begin{figure}
\includegraphics*[width=7.0cm,height=8.0cm,angle=-90]{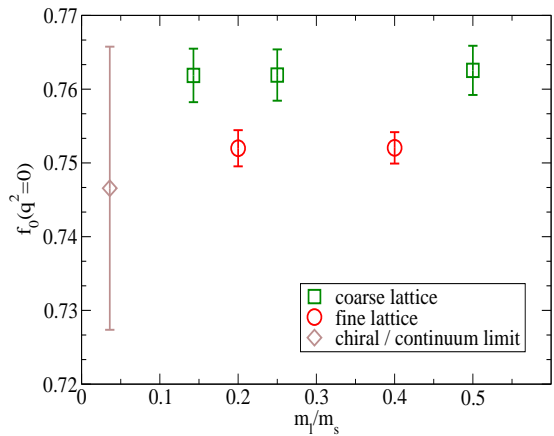}
\caption{ 
$f_0$ at $q^2 = 0$ for the five ensembles and in the physical limit.
 }
\label{f0_q2}
\end{figure}

\begin{figure}
\includegraphics*[width=7.0cm,height=8.0cm,angle=-90]{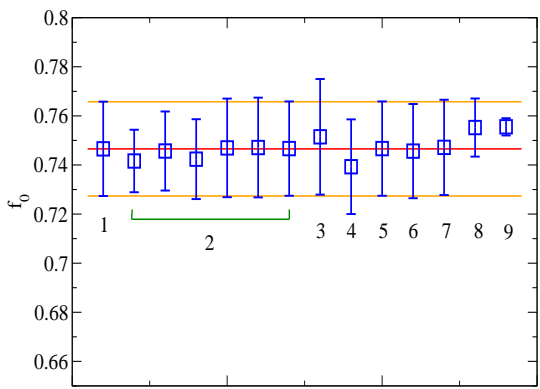}
\caption{
Tests of the ``simultaneous modified $z$-expansion extrapolation''. The red horizontal line is the central value of the fit shown in Figs.~\ref{coarse1} and \ref{fine1}, and the orange lines indicate the error.
The numbers under the data points correspond to the ``test numbers'' given in the text.
 }
\label{sys1}
\end{figure}

Having verified the efficacy of the $z$-expansion in fits to individual ensembles 
 we turn next to 
 modifying the fit ansatz to enable extrapolation to the physical limit.
All kinematic properties that depend on $q^2$ are absorbed by $P, \Phi_0,$ and $z$.
A natural way to distinguish between ensembles is to let $a_k \rightarrow a_k * D_k$, where $D_k$ contains the light quark mass and lattice spacing dependence as shown below (we set $k_{max} = 2$ and 
$P(q^2) = (1 - q^2/(M_{D_{s0}^*})^2)$).     
\begin{eqnarray}
\label{f0ansatz}
f_0(q^2) &=& \frac{1}{P(q^2) \, \Phi_0} \left (a_0 D_0 + a_1 D_1 z + a_2 D_2 z^2 \right )\nl 
& & \times (1+ b_1 (aE_K)^2 + b_2 (aE_K)^4),
\end{eqnarray}
where,
\begin{eqnarray}
\label{di}
D_i &=& 1 + c^i_1 x_l + c^i_2 \delta x_s +c^i_3 x_llog(x_l) + d_i (am_c)^2 \nl
   &&  + e_i (am_c)^4 + f_i \left ( \frac{1}{2} \delta M_\pi^2 + \delta M_K^2 \right ) \\
x_l &=& \frac{M_\pi^2}{(4 \pi f_\pi)^2} \\
\delta x_s &=& \frac{M_{\eta_s}^2-M_{\eta_s^{phys}}^2}{(4 \pi f_\pi)^2} \\
\delta M_\pi^2 &=& \frac{1}{(4 \pi f_\pi)^2} \left ( (M_\pi^{AsqTad})^2 - (M_\pi^{HISQ})^2
 \right )  \\
\label{lastparam}
\delta M_K^2 &=& \frac{1}{(4 \pi f_\pi)^2} \left ( (M_K^{AsqTad})^2 - (M_K^{HISQ})^2 \right). 
\end{eqnarray}
In eq. 34, we put typical analytic terms for light valence ($x_l$ and $\delta x_s$ terms) and sea quark mass ($\delta M_\pi$ and $\delta M_K$ terms) dependence.
We quote $M_K^{AsqTad}$ and $M_\pi^{AsqTad}$ from ref.~\cite{asqtad.meson}.
For the chiral logs, we only include up/down quark contributions.
The strange quark chiral logs are close to a constant that can be absorbed into the $a_i$'s.
There are two distinct sources of lattice spacing dependence.
$(am_c)^2$ and $(am_c)^4$ terms are due to the heavy quark discretization error, and $(aE_K)^2$ and $(aE_K)^4$ terms are introduced to estimate the discretization errors due to finite momentum.
Since we want the $a_i D_i$ to be independent of the momentum, the $aE_K$ terms are placed separately outside the $z$-expansion.
We include lattice spacing dependent terms up to fourth power, however we tested with even higher terms and confirmed that the higher terms are negligible (see Fig.~\ref{sys1}).
We have carried out simultaneous fits to all the data of 
Table~\ref{T.f0} using the above ansatz and find 
that very good fits are possible.   Figs.~\ref{coarse1} and \ref{fine1} 
plot the resulting fit curves for each ensemble and the 
chiral/continuum extrapolated 
curve with its error band for $f_0(q^2)$ versus $E_K^2$
(we show separately the coarse and fine ensembles in order 
to avoid too much clutter).  The fit is excellent and  has $\chi^2/dof = 0.44$.
In Table~\ref{T.aD}  we summarize fit results for $a_k * D_k$, k = 0,1,2 coming from the simultaneous fit 
both for individual ensembles and in the physical limit.  These are plotted in 
Figs.~\ref{a0} - \ref{a2}.  In Fig.~\ref{f0_q2} we plot 
$f_0(q^2=0)$ for the five ensembles and in the physical limit. One sees that
 within errors  this 
quantity shows little light quark mass dependence and a $\sim 1.3 \%$ lattice
 spacing dependence.

\vspace{.05in}
We call the chiral/continuum extrapolation based on the ansatz (\ref{f0ansatz})
- (\ref{lastparam}) and 
shown in Figs.~\ref{coarse1}, \ref{fine1} and \ref{f0_q2} the ``simultaneous modified $z$-expansion extrapolation.''  We have 
tested the stability of this extrapolation by adding further terms to the 
ansatz and/or modifying some of the fit parameters 
 and checking for changes in the physical limit $f_0(q^2=0)$.  
For example, we have,
\begin{enumerate}
\item added $x_l^2$  terms,
\item modified the lattice spacing dependent terms:
    \begin{enumerate}
    \item drop $(am_c)^4$ and $(aE_K)^4$ terms
    \item drop $(am_c)^4$ term
    \item drop $(aE_K)^4$ term
    \item add $(am_c)^6$ term
    \item add up to $(am_c)^{10}$ terms
    \item add $(aE_K)^6$ term,
    \end{enumerate}
\item  used $P(q^2) = 1$  for the pole term, 
\item used $E_K$ from dispersion relations,
\item used an overall factor $(1 + f (\frac{1}{2} \delta M_\pi + \delta M_K))$ 
outside the $\sum_k$ to incorporate sea quark effects, rather than include them in the $D_i$'s,
\item used an overall factor $(1 + d (am_c)^2 + e (am_c)^4)$ 
outside the $\sum_k$ to estimate the $am_c$ errors, rather than include them in the $D_i$'s,
\item replaced $\delta M_{\pi/K}^2 \rightarrow (M_{\pi/K}^{AsqTad})^2 / (4 \pi f_\pi)^2$,
\item used a simpler $D_i = 1 + d^i_1x_l + d^i_2 x_s$, 
\item used an even simpler $D_i = 1 + d^i_1x_l $.
\end{enumerate}
Fig.~\ref{sys1}  summarizes the results of these tests.
One sees that the standard $z$-expansion extrapolation
result is very robust.
The second item of the tests checks that we estimate the lattice spacing extrapolation error correctly. 
Until $(am_c)^4$ and $(aE_K)^4$ terms are included, the error is increasing, however after including the fourth powers the error is stabilized. 
This also shows that the $am_c$ error of the HISQ action is under control in our simulations.

\section{Results in the Physical Limit: $f_+(0)$, $|V_{cs}|$ and Unitarity Tests}
This section summarizes the main results of this paper.  We present our
Standard Model prediction for the
$D \rightarrow K, l \nu$ decay form factor at $q^2 = 0$,
$f_+(0) = f_0(0)$, determine the CKM matrix element $|V_{cs}|$ using
input from  BaBar and CLEO-c and carry out unitarity tests.

\vspace{.1in}
\noindent
\underline{$f_+(0)=f_0(0)$}

\vspace{.05in}
\noindent
We have seen that the simultaneous modified $z$-expansion extrapolation method gives very stable results.
It gives $f_+(0) = 0.748 \pm 0.019$ in the physical limit for $D^0 
\rightarrow K^- l\nu$, and $f_+(0) = 0.746 \pm 0.019$ for $D^+ \rightarrow \overline{K}^0 l\nu$.
We take an average over these two channels and our final result in the physical limit becomes,
\be
\label{resf+}
f^{D \rightarrow K}_+(0) = 0.747 \pm 0.011 \pm 0.015.
\ee
The first error comes from statistics and the second error represents systematic errors.
Table~\ref{T.error} summarizes the error budget.
One sees that the largest contributions to the total error come from statistics
followed by $(am_c)$ and $(aE_K)$ extrapolation errors.

In order to calculate the form factor, we have to put in meson masses from experiment and also from our lattice simulations. 
For example, we need experimental $D$, $K$, and $\pi$ meson masses to get the form factor at the physical limit, and $E_K$, $D$, and $K$ meson masses from the lattice calculations are used to fit at non-zero lattice spacing.
In Table~\ref{T.error}, ``Input meson mass'' refers to errors induced from these input meson masses.
In the fit ansatz, eq.~\ref{di}, there are light quark ($c_1^i$ and $c_3^i$), strange quark ($c_2^i$), and sea quark dependent terms ($f_i$). 
Each systematic error due to these terms is shown on the fourth to sixth line in the table.
Lattice spacing dependence errors are estimated separately for $(am_c)^n$ and $(aE_K)^j$ type contributions.

In the fit ansatz, $x_l log(x_l)$ is the most infrared sensitive term.
We calculate the pion-tadpole loop integral both at finite volume and at infinite volume and compare these to estimate the finite volume effects.
For the charm quark mass tuning error, we calculate the form factor with a different charm quark mass, $am_c = 0.629$, on the C3 ensemble, and compare with the result with the tuned $am_c = 0.6235$ of Table~\ref{T.qmass}.
All but the last two entries in Table~\ref{T.error} (finite volume and charm mass tuning)
were calculated using methods introduced in reference \cite{alpha}.  
The total error coming out
of the chiral/continuum extrapolation can be decomposed into individual
contributions, $\sigma^2 = \sum_i c_i \sigma^2_i$, where the sum $\sum_i$ goes
over the first 8 entries in Table~\ref{T.error}.  Details are described in Appendix B.

One might worry about other potential systematic errors, not listed in Table~\ref{T.error}, such as
those due to missing sea charm quarks or electromagnetism/isospin breaking.  The separate numbers given
above eq.~(\ref{resf+}) for $D^0 \rightarrow K^-$ and $D^+ \rightarrow \overline{K}^0$
form factors take into account just the differences in masses of the charged versus neutral
mesons.  This ``kinematic'' effect is seen to be less than $\sim 0.3$\%.  It is much harder to assess the
true dynamical electromagnetic effects.  However no statistically significant
differences have been observed experimentally \cite{cleo} and we will ignore
further electromagnetic/isospin breaking effects.  Similarly we will assume that errors due to missing
sea charm quarks are small enough so that they do not change the 2.5\% total error
when added in quadrature.  This has been true in the case of several quantities where
it was possible and appropriate to apply perturbative estimates of dynamical charm quark effects~\cite{fds2}.

\vspace{.1in}
\begin{table}
\caption{ Total error budget. \\
    }
\label{T.error}
\begin{center}
\begin{tabular}{|c|c|}
\hline
Type & Error \\
\hline
\hline
Statistical & 1.5 \% \\
Lattice scale ($r_1$ and $r_1/a$) & 0.2 \% \\
Input meson mass & 0.1 \% \\
Light quark dependence & 0.6 \% \\
Strange quark dependence & 0.7 \% \\
Sea quark dependence & 0.4 \% \\
$am_c$ extrapolation& 1.4 \% \\
$aE_K$ extrapolation& 1.0 \% \\
Finite volume & 0.01 \% \\
Charm quark tuning & 0.05 \% \\
\hline
Total & 2.5 \% \\
\hline
\end{tabular}
\end{center}
\end{table}

The total error for $f_+(0)$ is estimated here to be 2.5\%.  This is a factor of four times smaller than in the previous lattice calculation of Ref.~\cite{fermimilc05}.
This was achievable because of applying several new methods and techniques that were described in the text.
We employ the HISQ action for both charm and light quark actions and a scalar current rather than the traditional vector current.
Because of these new methods, we obtain results with smaller discretization errors and no operator matching.
We also developed the modified $z$-expansion extrapolation method, which is crucial to decrease errors due to the discretization, chiral / continuum extrapolation and parameterization of the form factor.
In order to decrease statistical errors, we apply random-wall sources and perform simultaneous fits with multiple correlators and $T$'s.
If we compare with the error budget of Ref.~\cite{fermimilc05}, then we see the statistical errors reduced from 3\% to 1.5\% and
the extrapolation and parameterization errors from 3\% to 1.5\% as well.
The biggest improvement is in the discretization errors. The total discretization errors have now been reduced from 9\% to 2\%.
We note that the concept of the discretization errors is different in Ref.~\cite{fermimilc05} compared to here.
In Ref.~\cite{fermimilc05}, they estimate the discretization errors by power counting, since they calculate at only one lattice spacing.
However we actually perform continuum extrapolations with correction terms for the discretization effects.
As a result, we do not have discretization errors per se, but instead extrapolation errors due to higher order correction terms.

\begin{figure}
\includegraphics*[width=7.0cm,height=8.0cm,angle=-90]{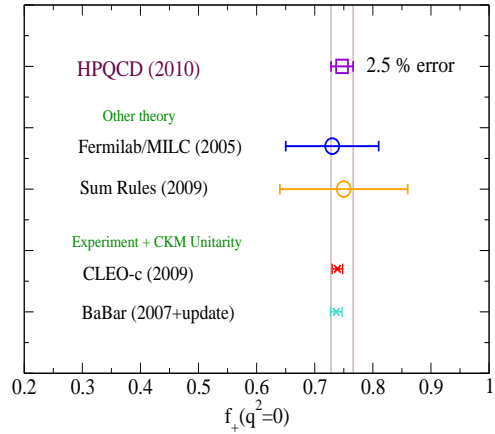}
\caption{ 
Comparisons of $f_0(q^2=0)$ with other calculations and experiments.
 }
\label{resultf0}
\end{figure}

\vspace{.1in}
In their papers both BaBar~\cite{babar} and CLEO-c~\cite{cleo} have converted their measurements of $f_+(0) * |V_{cs}|$ into results
for $f_+(0)$ using values for $|V_{cs}|$ fixed by CKM unitarity.  For this CLEO-c  uses 
the 2008 PDG CKM unitarity value of $|V_{cs}| = 0.97334(23)$ \cite{pdg2008} and obtains
$f^{D \rightarrow K}_+(0) = 0.739(9)$ 
 and BaBar uses $|V_{cs}| = 0.9729(3)$ leading to $f_+(0) = 0.737(10)$.
In Fig.~\ref{resultf0} we plot the new HPQCD result of this article, eq.(\ref{resf+}), together with earlier
theory results from the lattice \cite{fermimilc05} and from a recent sum rules calculation
\cite{sumrules} and with the BaBar and CLEO-c numbers. 
 One sees the very welcome reduction in theory errors which are now small enough so that the agreement between theory and experiment
in Fig.~\ref{resultf0} already provides a nontrivial indirect test of CKM unitarity.
  We can, however, do better and
carry out more direct tests of unitarity by  determining $|V_{cs}|$
without  the assumption of unitarity.

\vspace{.1in}
\noindent
\underline{Direct Determination of $|V_{cs}|$}

\begin{figure}
\includegraphics*[width=7.0cm,height=8.0cm,angle=-90]{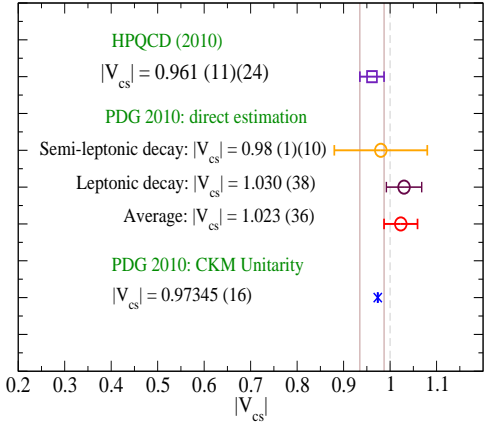}
\caption{
Comparisons of our new $|V_{cs}|$ with values in the PDG \cite{pdg}.
 }
\label{Vcs}
\end{figure}

\vspace{.05in}
\noindent
As experimental input we take
$f_+(0) * |V_{cs}| = 0.719 (8)$ from CLEO-c \cite{cleo} and
$f_+(0) * |V_{cs}| = 0.717 (10)$ from BaBar \cite{babar}.  For the latter we have multiplied
BaBar's quoted $f_+(0)$ with their quoted CKM unitarity value for $|V_{cs}|$.
Averaging between the two experiments we use
$f_+(0) * |V_{cs}| = 0.718 (8)$ together with eq.(\ref{resf+}) to extract $|V_{cs}|$.
One finds,
\be
\label{resvcs}
|V_{cs}| = 0.961 \pm 0.011 \pm 0.024,
\ee
in good agreement (as expected from Fig.~\ref{resultf0}) with the CKM unitarity value of 0.97345(16) \cite{pdg}.
The first error in (\ref{resvcs}) is from experiment and the second from the lattice calculation of this article.
This is a very precise direct determination of $|V_{cs}|$,
made possible by the many advances in lattice QCD that are described in this article
together with the tremendous progress in recent experimental studies of $D$ semileptonic decays
\cite{babar,cleo}.
In Fig.~\ref{Vcs} we plot several previous direct determinations of $|V_{cs}|$ from the 2010 PDG \cite{pdg}
together with (\ref{resvcs}) and the CKM unitarity value.

\vspace{.05in}
In a companion paper~\cite{fds2} where we update HPQCD's $D_s$ meson decay constant 
$f_{D_S}$, we also determine $|V_{cs}|$ from $D_s \rightarrow \tau \nu$ and 
$D_s \rightarrow \mu \nu$ leptonic decays.
One finds $|V_{cs}|_{leptonic} = 1.010 (22)$ which is consistent with eq.~(\ref{resvcs}) at the $1.4\; \sigma$ level.

\vspace{.1in}
\noindent
\underline{Further Unitarity Tests}

\begin{figure}
\includegraphics*[width=7.0cm,height=8.0cm,angle=-90]{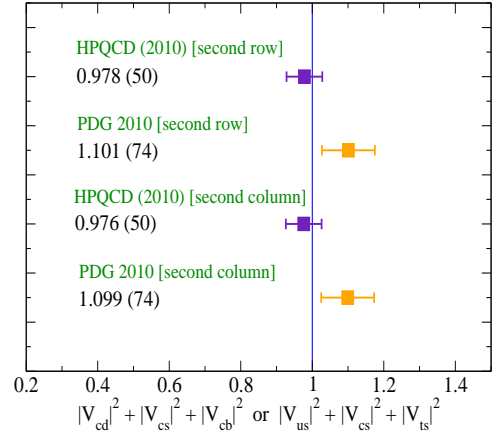}
\caption{
Unitarity checks for the second row and second column of the CKM matrix.
 }
\label{unitarity}
\end{figure}

\vspace{.05in}
\noindent
Using the new value of $|V_{cs}|$, eq.(\ref{resvcs}), and the current PDG  values
$|V_{cd}| = 0.230(11)$ and $|V_{cb}| = 0.0406(13)$ one finds,
\be
\label{2ndrow}
|V_{cd}|^2 + |V_{cs}|^2 + |V_{cb}|^2 = 0.978(50)
\ee
for the 2nd row. And similarly for the 2nd column, with $|V_{us}| = 0.2252(9)$ and $|V_{ts}= 0.0387(21)$
one gets,
\be
\label{2ndcol}
|V_{us}|^2 + |V_{cs}|^2 + |V_{ts}|^2 = 0.976 (50)
\ee
These 2nd row and 2nd column unitarity test results are shown in Fig.~\ref{unitarity} together
with the PDG numbers mentioned already in the Introduction.
Again one sees the improvement coming from the
reduction in the uncertainty in $|V_{cs}|$.

\section{Further Results from Two-point Correlators}
In this section we summarize 
physics results extracted from two-point correlators that emerged as part 
of our analysis of $D$ meson semileptonic decays.  This  includes 
determinations of decay constants, $f_\pi$, $f_K$, $f_D$ and $f_{D_s}$.  
These determinations serve mainly as consistency checks on our 
quark mass tunings and on our fitting and chiral/continuum extrapolation 
methods. More extensive studies of $f_{D_s}$ involving five lattice 
spacings are reported in \cite{fds2}.  
Here we present the first 
results for $f_D$ using HISQ charm and light quarks that employs the 
new HPQCD $r_1$ scale~\cite{r1} (the scale we use throughout in this article).  

\vspace{.05in}
We use
continuum partially quenched ChPT formulas augmented by discretization terms to 
extrapolate to the chiral/continuum limit. Error budgets are close to those in ref~\cite{hpqcdfds} with, however, a decrease in the $r_1$ uncertainty.
For $f_\pi$ and $f_K$, we find,
\be
f_\pi = 132.3 \pm 1.6 \, \mathrm{MeV}
\ee
\be
f_{K} = 157.9 \pm 1.5 \, \mathrm{MeV}
\ee
One sees that agreement with experiment~\cite{pdg} is good and  within  $1 \sim 1.5 \sigma$ (or 1.5\%).  

\begin{figure}
\includegraphics*[width=7.0cm,height=8.0cm,angle=-90]{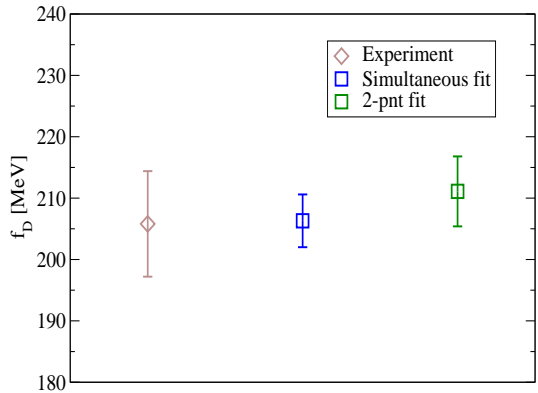}
\caption{ 
Comparisons of $f_D$ with experiment, simultaneous fit, and  two-point correlator fit.
 }
\label{fdd}
\end{figure}

\vspace{.05in}
For $f_D$ we have results from the simultaneous fit with three-point correlators 
and also from just the two-point correlator fits.
 As we noted in section V, the former has smaller errors however they are consistent with each other. 
\be
\label{fDsim}
f_D^{\mathrm{SimFit}} = 206.3 \pm 4.3 \, \mathrm{MeV}
\ee
\be
f_{D}^{\mathrm{2pt}} = 211.1 \pm 5.7 \, \mathrm{MeV}.
\ee
Both values show good agreement with experiment~\cite{pdg} as can be seen in Fig.~\ref{fdd}.

For $f_{D_s}$, we find,
\be
\label{ourfds}
f_{D_s} = 250.2 \pm 3.6 \ \mathrm{MeV},
\ee
which agrees with (but is less precise than) our recent update in reference~\cite{fds2}, $f_{D_s} = 248.0 \pm 2.5$ MeV.
One sees that with experimental values
having come down in recent years and with the increase in the HPQCD value, there is
no longer any discrepancy (beyond 1.6~$\sigma$) between theory and experiment.
The current HFAG~\cite{hfag} number is $f_{D_s} = 257.3 \pm 5.3 \, \mathrm{MeV}$.

\begin{figure}
\includegraphics*[width=7.0cm,height=8.0cm,angle=-90]{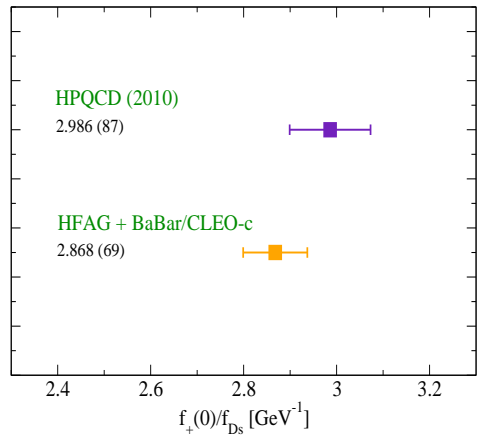}
\caption{
Comparisons of $f^{D \rightarrow K}_+(0)/f_{D_s}$. We estimate the experiment value from a simple calculation of combining $f_{D_s}$ from HFAG~\cite{hfag} and 
$f^{D \rightarrow K}_+(0)$ from BaBar~\cite{babar} and CLEO-c~\cite{cleo}.
This is an appropriate thing to do since both quantities were derived from
experimental measurements using the unitarity value of $|V_{cs}|$.}
\label{ratiofpfds}
\end{figure}

Finally we present the ratio,
\be
\label{ratio.f0fds}
\frac{f^{D \rightarrow K}_+(0)}{f_{D_s}} = 2.986 \pm 0.087\, \mathrm{GeV}^{-1}.
\ee
This quantity can also be obtained from experimental measurements of 
$ D \rightarrow K, l \nu$ semileptonic and $D_s$ leptonic decay branching 
fractions, and has the virtue that $|V_{cs}|$ drops out in the 
ratio.
We compare (\ref{ratio.f0fds}) with experiment in Fig.~\ref{ratiofpfds} and good agreement is found.

\section{Summary and Future Outlook}
We have completed the first study of $D \rightarrow K$  semileptonic decays using the HISQ
action for the valence charm, strange and light quarks. The most important result
 of this article  is given in eq.~(\ref{resf+})
 and provides the form factor $f^{D \rightarrow K}_+(q^2)$ at $q^2 = 0$.
 We were able to determine this quantity
with a 2.5\% total error which represents a four fold improvement in precision over earlier
theory results.  This is shown in Fig.~\ref{resultf0}.
We then combined our form factor result with recent measurements
of $D \rightarrow K$ semileptonic decays by the BaBar and CLEO-c collaborations to
extract a very accurate direct determination of the CKM matrix element $|V_{cs}|$.
This is given in eq.~(\ref{resvcs}) and comparisons with previous determinations shown in
Fig.~\ref{Vcs}. The new value for $|V_{cs}|$ is consistent
 with the PDG value based on CKM unitarity.
We carried out direct tests of 2nd row and 2nd column unitarity with results given in
equations (\ref{2ndrow}) and (\ref{2ndcol}) and depicted in Fig.~\ref{unitarity}.
  Although still far from the accuracy achieved in
 1st row unitarity tests, the reduction in errors on $|V_{cs}|$ has made 2nd row and column
unitarity tests much more relevant and interesting than in the past.
In section VIII we give a lattice QCD value for $f^{D \rightarrow K}_+(0)/f_{D_s}$
which is consistent with experiment.
Within the current theory and experimental errors, this provides a highly nontrivial
consistency check on how we treat $D$ semileptonic and leptonic decays on the
lattice and more generally in the Standard Model.

\vspace{.2in}
The calculations of this article can be improved upon and extended in several ways.
The largest errors in Table~\ref{T.error} from  statistics and continuum extrapolations can be
reduced straightforwardly by working with more gauge configurations and time sources
and at more values of the lattice spacing. An obvious extension of the current
study  is to investigate
$D \rightarrow \pi$ semileptonic decays and determine $|V_{cd}|$.  Work on
this project has already begun.  In a future project we also plan to calculate
the vector current hadronic matrix elements $\langle K(\pi)|V_\mu| D \rangle$.
This will provide  $f_+(q^2)$ as a function of $q^2$.  As mentioned in section II
we will be carrying out nonperturbative matching of the vector current
based on PCVC for this project.

\vspace{.3in}
\underline{{\bf Acknowledgements}}\\
We thank the MILC collaboration for making their AsqTad $N_f = 2 +1 $ configurations
available.  This work was supported in part by the DOE and NSF in the US and by STFC
in the UK.  Computations were carried out at the Ohio Supercomputer Center and
on facilities of the USQCD collaboration funded by the Office of Science of the
U.S. DOE.

\appendix
\section{Priors and Prior Widths for Section V}
We list a sample set of priors and prior widths in Table~\ref{T.prior} that have been used for the simultaneous fits to $C_D^{2pnt}$, $C_K^{2pnt}$, and three of the three-point correlators with $T = 19$, $20$, and $23$
for ensemble F1. 
The fit ansatz and results have been presented in Section V.
\begin{table}
\caption{
A sample set of the priors and prior widths for ensemble F1 with $\vec{p} = (0,0,0)$ and $\vec{p} = (1,1,0)$. 
We have tested with various priors and prior widths, and the fit results are not sensitive to reasonable variations. 
Note that $i = 1, 2, 3, ...$ and $j,k = 0, 1, 2, ...$.
}
\begin{center}
\label{T.prior}
\begin{tabular}{|c|cc|cc|}
\hline
 & prior  &  width  &  prior  & width\\
 &$ \vec{p} = (0,0,0)$ && $ \vec{p} = (1,1,0)$& \\
\hline
\hline
  $A_{jk}$&0.01  & 0.1 & 0.01 & 0.1 \\
  $B_{jk}$&0.01  & 0.1 & 0.01 & 0.1 \\
  $C_{jk}$&0.01  & 0.1 & 0.01 & 0.1 \\
  $D_{jk}$&0.01  & 0.1 & 0.01 & 0.1 \\
\hline
$E_0^D$ & 0.815 & $^{+0.815}_{-0.408}$ & 0.8 & $^{+0.8}_{-0.4}$ \\
$E_i^D-E_{i-1}^D$ & 0.6 & $^{+0.6}_{-0.3}$ & 0.4 & $^{+0.4}_{-0.2}$ \\
$b_0^D$ & 0.12 & 1.0 & 0.015 & 0.3 \\
$b_i^D$ & 0.1 & 1.0 & 0.03 & 0.3 \\
\hline
$E_0^{\prime D}$ & 1.0 & $^{+1.0}_{-0.5}$ & 1.0 & $^{+1.0}_{-0.5}$ \\
$E_i^{\prime D}-E_{i-1}^{\prime D}$ & 0.4 & $^{+0.4}_{-0.2}$ & 0.4 & $^{+0.4}_{-0.2}$ \\ 
$d_0^D$ & 0.01 & 0.1 & 0.0028 & 0.1 \\
$d_i^D$ & 0.01 & 0.1 & 0.006 & 0.1 \\
\hline
$E_0^K$ & 0.23 & $^{+0.23}_{-0.12}$ & 0.39 & $^{+0.39}_{-0.2}$ \\
$E_i^K-E_{i-1}^K$ & 0.5 & $^{+0.5}_{-0.25}$ & 0.4 & $^{+0.4}_{-0.2}$ \\         
$b_j^K$ & 0.15 & 1.0 & 0.01 & 0.1 \\
\hline
$E_0^{\prime K}$ & 0.4 & $^{+0.4}_{-0.2}$ & 0.53 & $^{+0.53}_{-0.27}$ \\
$E_i^{\prime K}-E_{i-1}^{\prime K}$ & 0.5 & $^{+0.5}_{-0.25}$ & 0.4 & $^{+0.4}_{-0.2}$ \\
$d_j^K$ & 0.01 & 0.1 & 0.001 &0.01  \\
\hline
\end{tabular}
\end{center}
\end{table}

\section{Bayesian Fits in Section VI}

From (\ref{f0ansatz}) - (\ref{lastparam}) one sees that the 
 chiral/continuum extrapolation ansatz for $f_0(q^2)$ starts out 
with 23 basic fit parameters $c_n$, n = 1,2,...,23.

\begin{equation}
c_n : \;
a_0,\; a_1,\; a_2,\; b_1,\; b_2,\; c_1^i,\; c_2^i,\; c_3^i,\; d_i,\; e_i,\; f_i,
\end{equation}
where $i = 0,1,$ and $2$.
In a Bayesian fit each of these fit parameters will have its own prior 
$\overline{c}_n$ and prior width $\sigma_n$, which we call ``Group I.''  Our choices for these priors 
will be discussed below.  Our fit ansatz for $f_0$ includes in addition to 
the fit parameters $c_n$ also many input parameters such as $M_D$, $E_K$, 
$r_1$ etc. all of which have  some uncertainty associated with them.  Ref. \cite{alpha}
describes a method to include effects coming from these types of uncertainties 
into  the final error in the extrapolated value for $f_0$.    What one does is 
convert all these input parameters into additional fit parameters with 
priors and prior widths given by their known central values and errors 
Using this 
approach we have changed 61 input parameters into new fit parameters $p_j$, 
j = 1,2,....,61, which we call ``Group II.''
\begin{eqnarray}
p_j : & &  \left (\frac{r_1}{a} \right )^i, aM_D^i, aE_K^i(\vec{p}), aM_{\eta_s}^i, aM_\pi^i,  \nl
 && (aM_K^{asqtad})^i, (aM_\pi^{asqtad})^i, M_{D_s^*}^i,  \nl
 &&   r_1, M_{\eta_s}^{phys}, M_\pi^{phys}, M_D^{phys}, M_K^{phys}, M_{D_s^*}^{phys}
\end{eqnarray}
where $i = 1,2,..,5$ goes over the 5 ensembles.  The use of 
Group II parameters is a very efficient way to include errors coming from 
uncertainties in input parameters into our final total error. 
An alternative approach would require visiting each input parameter in turn, 
redoing the chiral/continuum extrapolation and coming up with an estimate for 
the systematic error coming from this input parameter. In our approach no 
additional systematic errors for these input parameters are  called for and 
the effects from their uncertainties are included in 
the extrapolation error.
{ For instance, this is very helpful to include the error from $E_K^2$.
In Fig.~\ref{coarse1} and \ref{fine1}, there are errors for $E_K^2$ on the lattice data as well as the extrapolated results, and estimating these errors is not a trivial task. 
However, introducing the Group II parameters incorporates all $E_K^2$ errors as part of the final vertical error.}

\vspace{.05in}
In Bayesian fits one minimizes the augmented chisquared,
\begin{eqnarray}
\chi^2_{aug} &=& \chi^2_{traditional} + \chi^2_{\mathrm{Group\;I}} + \chi^2_{\mathrm{Group\; II}} \nl
 && \\
\chi^2_{traditional} & = & \sum_{i = 1}^{20} \frac{(f_0^i - f_0(ansatz))^2}
{(\sigma^i_{f_0})^2}  \\
 & & \nl
\chi^2_{\mathrm{Group\; I}} & = & \sum_{n = 1}^{23} \frac{(c_n - \overline{c}_n)^2}
{\sigma^2_n}  \\
 & & \nl
\chi^2_{\mathrm{Group\; II}} & = & \sum_{j = 1}^{61} \frac{(p_j - \overline{p}_j)^2}
{\sigma^2_j}  
\end{eqnarray}

When carrying out the chiral/continuum extrapolations we have expressed all 
dimensionful quantities in units of GeV. 
We give the set of priors and prior widths for the Group I parameters 
$c_n$ in Table~\ref{T.group1} and for the Group II parameters $p_j$ in Table~\ref{T.group2_1} and \ref{T.group2_2}.

\begin{table}
\caption{
Priors and prior width of the Group I parameters for the simultaneous modified $z$-expansion extrapolation fit
}
\label{T.group1}
\begin{center}
\begin{tabular}{|c|cc|cc|}
\hline
Group I & prior  &  width  &  fit result  &fit error \\
\hline
\hline
 $a_0$ & 0&1&  0.09766& 0.0029 \\
 $a_1$ & 0&1& 0.08999&  0.02\\
 $a_2$ & 0&1& -0.07044& 0.11 \\
  \hline
  $b_1$ &0 &0.3&0.03775& 0.13 \\
  $b_2$ & 0&0.3&0.07179& 0.17 \\
  \hline
 $c_1^0$   &0 &1& -0.52596 & 0.31 \\
 $c_1^1$ & 0&1& -0.19051 & 0.82 \\
 $c_1^2$ &0 &1& 0.02877 & 1 \\
  \hline
  $c_2^0$  &0 &1& -0.09919 & 0.98 \\
 $c_2^1$ & 0&1& 0.00827 & 1 \\
 $c_2^2$ & 0&1& 0.00044 & 1 \\
  \hline
  $c_3^0$ & 0&1& -0.02897 & 0.24 \\
  $c_3^1$& 0&1& 0.32804 & 0.66 \\
  $c_3^2$& 0&1&  -0.03116& 1 \\
  \hline
  $d_0$& 0&0.3&  0.00966& 0.11 \\
  $d_1$& 0&0.3&  0.01769& 0.29 \\
  $d_2$& 0&0.3&  0.00541& 0.3 \\
  \hline
  $e_0$& 0&0.2&  0.01554& 0.19 \\
  $e_1$& 0&0.2&  0.00447&  0.2\\
  $e_2$& 0&0.2&  0.00131&  0.2\\
  \hline
  $f_0$& 0&0.3&  -0.10860& 0.28 \\
  $f_1$& 0&0.3&  -0.00474& 0.3 \\
  $f_2$& 0&0.3&  0.00105& 0.3 \\
\hline
\end{tabular}
\end{center}
\end{table}

\begin{table}
\caption{
Priors and prior width of the Group II parameters for the simultaneous modified $z$-expansion extrapolation fit. Parameters with five rows correspond to that on the five ensembles, C1, C2, C3, F1, and F2.
}
\label{T.group2_1}
\begin{center}
\begin{tabular}{|c|cc|cc|}
\hline
Group II & prior  &  width  &  fit result  &fit error \\
\hline
\hline
 $r_1$ &0.3133 & 0.0023 & 0.313285 & 0.0023 \\
 $M_{\eta_s}^{phys}$ & 0.6858 & 0.004 & 0.685799 & 0.004 \\
 $M_\pi^{phys}$ & 0.1373 & 0.0023 & 0.1373 & 0.0023 \\
 $M_D^{phys}$ & 1.8645 & 0.0004 & 1.8645 & 0.0004 \\
 $M_K^{phys}$ & 0.4937 & 0.000016 & 0.4937 & 0.000016 \\
 $M_{D_s^*}^{phys}$ & 2.3173 & 0.0006 & 2.3173 & 0.0006 \\
  \hline
  $r_1/a$& 2.647 & 0.003 & 2.64677 & 0.003 \\
  & 2.618 & 0.003 & 2.61818 & 0.003 \\
  & 2.644 & 0.003 & 2.64388 & 0.003 \\
  & 3.699 & 0.003 & 3.69905 & 0.003 \\
  & 3.712 & 0.004 & 3.71213 & 0.0039 \\
  \hline
  $aM_D$ & 1.13927 & 0.00066 & 1.13925 & 0.00066 \\
  & 1.15947 & 0.00076 & 1.15949 & 0.00076 \\
  & 1.16179 & 0.0005 & 1.16179 & 0.0005 \\
  & 0.814062 & 0.00035 & 0.814066 & 0.00035 \\
  & 0.819663 & 0.00026 & 0.819663 & 0.00026 \\
  \hline
  $aE_K$ & 0.312174 & 0.0002 & 0.312159 & 0.0002 \\
  $\vec{p}=(0,0,0)$& 0.32851 & 0.00048 & 0.328529 & 0.00048 \\
  & 0.357205 & 0.00022 & 0.357228 & 0.00022 \\
  & 0.228546 & 0.00017 & 0.228572 & 0.00017 \\
  & 0.24596 & 0.00014 & 0.245944 & 0.00014 \\
  \hline
  $aE_K$& 0.408141 & 0.00066 & 0.408059 & 0.00065 \\
  $\vec{p}=(1,0,0)$& 0.453061 & 0.0016 & 0.453098 & 0.0015 \\
  & 0.475036 & 0.00087 &0.474849  & 0.00086 \\
  & 0.320348 & 0.00069 & 0.320396 & 0.00067 \\
  &  0.333981& 0.00037 & 0.333932 & 0.00036 \\
  \hline
  $aE_K$& 0.483702 & 0.00094 & 0.483798 & 0.00093 \\
  $\vec{p}=(1,1,0)$& 0.552504 & 0.0017 & 0.552527 & 0.0016 \\
  & 0.572004 & 0.001 & 0.572016 & 0.001 \\
  & 0.391932 & 0.00087 & 0.391883 & 0.00085 \\
  & 0.401385 & 0.00065 & 0.401619 &  0.00064\\
  \hline
  $aE_K$& 0.546941 & 0.002 & 0.547535 & 0.002 \\
  $\vec{p}=(1,1,1)$& 0.637288 & 0.0032 & 0.637096 & 0.0032 \\
  & 0.652424 & 0.0022 & 0.652314 & 0.0022 \\
  & 0.45593 & 0.0015 & 0.455311 & 0.0015 \\
  & 0.460897 & 0.0011 & 0.460966 & 0.0011 \\
  \hline
  $aM_{\eta_s}$& 0.411128 & 0.00018 & 0.41113 & 0.00018 \\
  & 0.414346 & 0.00022 & 0.414345 & 0.00022 \\
  & 0.411848 & 0.00022 & 0.411847 & 0.00022 \\
  & 0.294159 & 0.00012 & 0.294158 & 0.00012 \\
  & 0.293114 & 0.00018 & 0.293117 &  0.00018\\
  \hline
  $aM_\pi$& 0.159893 & 0.00017 & 0.159895 & 0.00017 \\
  & 0.210815 & 0.00023 & 0.210813 & 0.00023 \\
  & 0.293124 & 0.00023 & 0.293119 & 0.00023 \\
  & 0.134449 & 0.00015 & 0.134445 & 0.00015 \\
  & 0.187346 & 0.00013 &  0.18735& 0.00013 \\
  \hline
\end{tabular}
\end{center}
\end{table}

\begin{table}
\caption{
(Continued) Priors and prior width of the Group II parameters for the simultaneous modified $z$-expansion extrapolation fit. Parameters with five rows correspond to that on the five ensembles, C1, C2, C3, F1, and F2.
We quote $M_K^{AsqTad}$ and $M_\pi^{AsqTad}$ from ref.~\cite{asqtad.meson}.
}
\label{T.group2_2}
\begin{center}
\begin{tabular}{|c|cc|cc|}
\hline
Group II & prior  &  width  &  fit result  &fit error \\
\hline
\hline
  $aM_K^{asqtad}$& 0.3653 & 0.00029 & 0.365304 & 0.00029 \\
  & 0.38331 & 0.00024 & 0.383309 & 0.00024 \\
  & 0.40984 & 0.00021 & 0.409839 & 0.00021 \\
  & 0.25318 & 0.00019 & 0.253177 & 0.00019 \\
  & 0.27217 & 0.00021 & 0.272174 & 0.00021 \\
  \hline
  $aM_\pi^{asqtad}$& 0.15971 &0.0002  & 0.15971 & 0.0002 \\
  & 0.22447 & 0.00017 & 0.22447 & 0.00017 \\
  & 0.31125 & 0.00016 & 0.31125 &  0.00016\\
  &  0.14789& 0.00018 & 0.147889 & 0.00018 \\
  & 0.20635 & 0.00018 & 0.206351 & 0.00018 \\
  \hline
  $M_{D_s^*}$& 2.32897 & 0.005 & 2.33033 & 0.0047 \\
  & 2.32899 & 0.005 & 2.32818 & 0.0048 \\
  & 2.33053 & 0.005 & 2.33028 & 0.0048 \\
  & 2.32293 & 0.005 & 2.32179 & 0.0045 \\
  & 2.32113 & 0.005 & 2.32197 & 0.0045 \\
\hline
\end{tabular}
\end{center}
\end{table}
Setting priors and prior widths is straightforward.
For Group I, the quark mass terms, such as $x_l$ and $x_s$, are normalized by the scale, $\Lambda \equiv 4\pi f_\pi$.
Therefore, it is natural that we expect that the parameters vary between $-1$ to $1$.
However, we know from other lattice calculations with the same gauge configurations and our lattice data that the sea quark mass contribution is smaller than that of valence quark.
Thus, we take the priors and prior widths for the sea quark mass terms as $f_i = 0 \pm 0.3$.
The leading heavy quark error is proportional to $\mathcal{O}(\alpha_s (am_c)^2)$
so we use $d_j = 0 \pm 0.3$ for the $(am_c)^2$ terms.
For the purposes of setting priors, we conservatively do not include a factor of
$v^2/c^2$ here.
On the other hand for the $(am_c)^4$ terms we do take the expected factor of
$v^2/c^2$ into account and choose $e_i = 0 \pm 0.2$.
Similarly we use $b_j = 0 \pm 0.3$ for the $(aE_K)^2$ and $(aE_K)^4$ terms.
This reflects a factor of $\alpha_s$ for the $(aE_K)^2$ terms and the fact that
higher powers of $(ap^\mu)$ typically come with smaller numerical factors relative to lower 
powers (such as in an expansion of $\frac{1}{a}sinh (aE)$).
For Group II, we use lattice results and experiments that we described in the text for the priors and prior widths. 

As we stated above, all sources of systematic errors are already included in the fit ansatz, except the finite volume and charm quark mass tuning errors.
We can consider the total error squared, $\sigma^2$, as a linear combination of each source \cite{alpha};
\begin{equation}
\sigma^2 =  \sum_{n=1}^{23} C_{c_n} \sigma_{c_n}^2 + \sum_{j=1}^{61} C_{p_j} \sigma_{p_j}^2,
\end{equation} 
where the first term is for Group I ($c_n$), and the second term is for Group II ($p_j$).
We actually calculate the contributions from each source, $C_{c_n} \sigma_{c_n}^2$ and $C_{p_j}\sigma_{p_j}^2$ using the method presented in \cite{alpha}, and they add up to the total error $\sigma^2$ correctly.
We group together appropriate parameters, and list them in Table~\ref{T.error}.

\section{Chiral and Continuum Extrapolations based on Chiral Perturbation Theory}

In this Appendix we carry out  further consistency tests of the chiral/continuum 
extrapolation of section VI by working with a completely independent 
fit ansatz. 
  We will use the partially quenched chiral perturbation 
theory (PQChPT) formulas developed in Refs.\cite{becirevic, bernard} augmented by 
terms parameterizing discretization effects and $E_K$ dependence.

\begin{figure}
\includegraphics*[width=7.0cm,height=8.0cm,angle=-90]{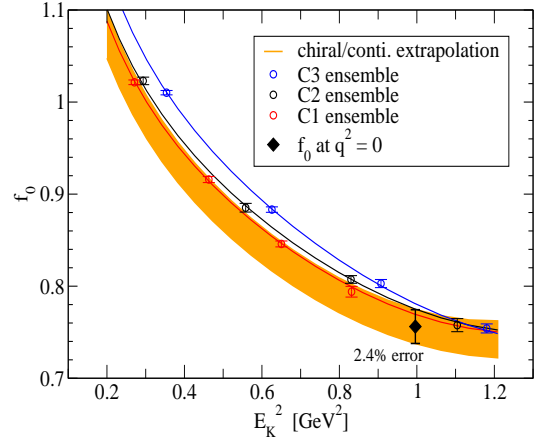}
\caption{
Chiral/continuum extrapolation of $f_0(q^2)$ versus $E_K^2$ from 
the ChPT ansatz.  The data points are coarse lattice points.  
Three individual curves and the extrapolated band are from a fit to all five ensembles.
 }
\label{coarse2}
\end{figure}

\begin{figure}
\includegraphics*[width=7.0cm,height=8.0cm,angle=-90]{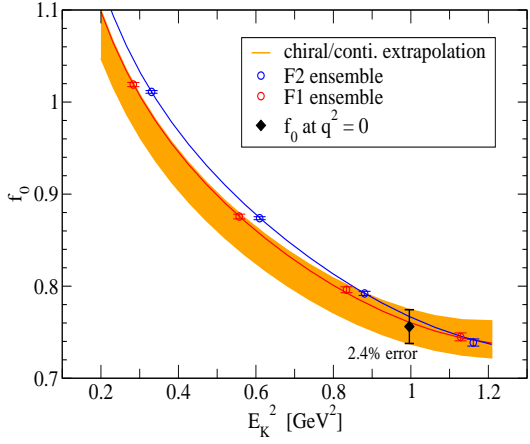}
\caption{
Chiral/continuum extrapolation of $f_0(q^2)$ versus $E_K^2$ from 
the ChPT ansatz.  The data points are fine lattice points.  
Two individual curves and the extrapolated band are from a fit to all five ensembles.
 }
\label{fine2}
\end{figure}

\vspace{.05in}
Heavy meson ChPT formulas are organized through form factors $f_\parallel$ 
and $f_\perp$ in terms of which $f_0(q^2)$ is given by,
\be
\label{f0comb}
f_0(q^2) = \frac{\sqrt{2 M_D}}{M_D^2 - M_K^2} \left [ (M_D - M_K) f_\parallel 
 + (E_K^2 - M_K^2) f_\perp \right ].
\ee
We follow very 
closely the approach and notation of Ref.\cite{bernard}, however with all the 
taste breaking effects turned off.
 $f_\parallel$ and $f_\perp$ are parameterized as,
\begin{eqnarray}
\label{fpara}
 f_\parallel &=&  
 \frac{\kappa}{f_\pi} \left [ 1 + \delta f_\parallel + c_l^\parallel m_l + 
c_{s^\prime}^\parallel m_{s^\prime}  + \right. \nl
& & \left. c^\parallel_{sea} ( 2 m_u + m_s) 
+ h_\parallel(E_K) \right ] ( 1 + c_0 (am_c)^2 \nl 
& & + c_1 (am_c)^4)  \\
\label{fperp}
 f_\perp &=&  
 \frac{\kappa}{f_\pi} \frac{g_\pi}{(E_k + \Delta^* + D)} \left [ 1 + \delta f_\perp 
+ c_l^\perp m_l +  \right. \nl
& &   (c_l^\parallel + c_l^\perp - c_{s^\prime}^\parallel) m_{s^\prime}  + 
  c^\perp_{sea} ( 2 m_u + m_s) + \nl
& & \left. h_\perp(E_K) \right ]
 ( 1 + c_0 (am_c)^2 + c_1 (am_c)^4).
\end{eqnarray}
The chiral logs are contained in $\delta f_\parallel$, $\delta f_\perp$
and $D$.  
We give their explicit expressions in Appendix D. 
$m_l$ and $m_{s^\prime}$ 
are valence and $m_u$ and $m_s$ the sea quark masses.
$g_\pi$ is the $D D^* \pi$ coupling and $\Delta^*$  the $D_s^* - D$ 
mass splitting. $h_\parallel(E_K)$ and $h_\perp(E_K)$ are unknown 
functions of $E_K$.  We will use polynomial expansions for them.  
$h_{\parallel,\perp}(E_K) = c_1^{\parallel,\perp} E_K + c_2^{\parallel,\perp} 
E_K^2 +  ...$.  The first two terms are motivated by ChPT \cite{bernard}, 
however we view $h_{\parallel,\perp}
(E_K)$ as potentially parameterizing  $f_0$ more generally, even beyond the regime of 
small $E_K$ where ChPT is assumed valid.  Our data is fit very well ($\chi^2/{dof}$=0.48), however, 
all the way to $E_K \approx 1$GeV keeping just 
terms through ${\cal O}(E_K^2)$ in $h_\parallel$ and $h_\perp$.
 Figs.~\ref{coarse2} and \ref{fine2} show results from a simultaneous fit to all our data points using 
the ChPT ansatz.  These should be compared to Figs.~\ref{coarse1} and \ref{fine1} from the modified $z$-expansion 
ansatz.

\begin{figure}
\includegraphics*[width=7.0cm,height=8.0cm,angle=-90]{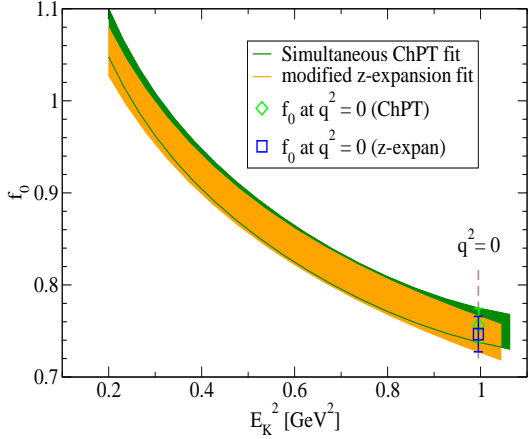}
\caption{
Comparisons of $f_0(q^2)$ in the physical limit from the $z$-expansion and 
the ChPT extrapolations.
 }
\label{f0_conti}
\end{figure}

\begin{figure}
\includegraphics*[width=7.0cm,height=8.0cm,angle=-90]{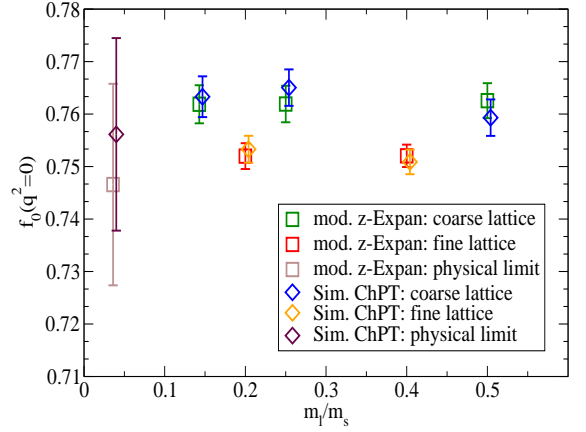}
\caption{
Comparisons of 
$f_0$ at $q^2 = 0$ for the five ensembles and in the physical limit 
from the $z$-expansion and the ChPT extrapolations.
 }
\label{f0_q22}
\end{figure}

\vspace{.05in}
In Fig.~\ref{f0_conti} we compare  $f_0(q^2)$ in the physical limit coming from 
the $z$-expansion extrapolation of section VI and the ChPT extrapolation of this Appendix 
over the entire physical $q^2$ range.  And in Fig.~\ref{f0_q22} we compare results  at $q^2=0$ 
for each ensemble and in the physical limit.  One sees that the two extrapolations 
are nicely consistent with each other.  
We believe the consistency check of this Appendix has been very useful.  
It provides further support for the results of the $z$-expansion extrapolation and 
 indicates that errors there were not underestimated.
Note that the ChPT ansatz includes all the complicated chiral logs of Appendix D,
whereas the $z$-expansion ansatz makes do with just a simple $x_l log x_l$ term.

\section{Partially Quenched ChPT Chiral Logs}
In this appendix we summarize partially quenched ChPT (PQChPT) expressions for the 
chiral logarithm terms $\delta f_\parallel$, $\delta f_\perp$ and $D$ that we 
employ in 
eqs.(\ref{fpara}) and (\ref{fperp}).  The formulas for PQChPT in continuum QCD 
for heavy-to-light semileptonic decays were first developed in Ref.\cite{becirevic}
 for degenerate 
sea quarks. Ref.\cite{bernard}
 generalized these results to nondegenerate 1+1+1 sea quarks and 
also to Staggered ChPT.  We have started from the 1+1+1 continuum PQChPT expressions 
given in Ref. \cite{bernard} for individual diagrams and for the $D$ and kaon wave function 
renormalizations to obtain the full $\delta f_{\parallel,\perp}$ in the 2+1 
PQChPT case.  For the convenience of the reader we give these reconstructed 
expressions below. We use the same notation as in Ref.\cite{bernard}. 
 ``$x$'' and ``$y$'' 
stand for the light valence quarks in the daughter meson
 (for the kaon $x \equiv l$ and 
$y \equiv s^\prime$) and ``$u$'' and ``$s$'' denote sea light and strange quarks.  
Furthermore $m_{ab}$ is the mass of the pseudoscalar meson with quark content $a$ and 
$b$ and $m^2_\eta = \frac{1}{3}(m^2_{uu} + 2 m^2_{ss})$. 
\begin{eqnarray}
\label{dfpara}
 & & (4 \pi f)^2 \delta f^{D \rightarrow K}_\parallel = \nl
& & \left \{ \left [ I_1(m_{yu}) + \frac{1}{2} I_1(m_{ys}) \right] - 3 g^2_\pi 
\left [ I_1(m_{xu}) + \frac{1}{2} I_1(m_{xs}) \right ] \right . \nl
& & + \left [ 2 I_2(m_{yu}) + I_2(m_{ys}) \right ] \nl
&& + \frac{1}{3} \left [ R_x^{[3,2]}(m_{xx}) \; (I_1(m_{xx}) + I_2(m_{xx}))
 \; + \right . \nl
&&  \qquad  R_y^{[3,2]}  (m_{yy}) \; (I_1(m_{yy}) + I_2(m_{yy}))  \; + \nl
&&  \left. \qquad  R_\eta^{[3,2]}(m_\eta) \; (I_1(m_\eta) + I_2(m_\eta)) \right ] \nl
& & + \frac{1}{6} \left [ DR^{[2,2]}(m_{yy}; I_1) \right ] 
 - \frac{3 g^2_\pi}{6} \left [ DR^{[2,2]}(m_{xx}; I_1) \right ] \nl
& & \left .+ \frac{1}{3} \left [ DR^{[2,2]}(m_{yy}; I_2) \right ]  \right \}
\end{eqnarray}
\begin{eqnarray}
\label{dfperp}
 & & (4 \pi f)^2 \delta f^{D \rightarrow K}_\perp = \nl
& & \left \{ - \left [ I_1(m_{yu}) + \frac{1}{2} I_1(m_{ys}) \right] - 3 g^2_\pi 
\left [ I_1(m_{xu}) + \frac{1}{2} I_1(m_{xs}) \right ] \right . \nl
&& - \frac{g^2_\pi}{3} \left [ R_x^{[3,2]}(m_{xx}) \; K_1(m_{xx}) + \right . \nl
&& \left . \qquad \;\;  R_y^{[3,2]}  (m_{yy}) K_1(m_{yy}) + 
 R_\eta^{[3,2]}(m_\eta) K_1(m_\eta) \right ] \nl
& & \left . - \frac{1}{6} \left [ DR^{[2,2]}(m_{yy}; I_1) \right ] 
 - \frac{3 g^2_\pi}{6} \left [ DR^{[2,2]}(m_{xx}; I_1) \right ] \right \} \nl
\end{eqnarray}
\begin{eqnarray}
\label{dd}
& &  (4 \pi f)^2 \, D^{D \rightarrow K}  =  
 -3 g^2_\pi (v \cdot p) \; \times \nl
&& \left \{ \left [ 2 K_1(m_{yu}) + K_1(m_{ys}) 
\right ]  + 
\frac{1}{3} \left [ DR^{[2,2]}(m_{yy}; K_1) \right ] \right \} \nl
\end{eqnarray}
In the $D$ meson restframe $v \cdot p = E_K$. Furthermore one has,
\be
I_1(m) = m^2 log\frac{m^2}{\Lambda^2}
\ee
\be
I_2(m) = - 2 (v \cdot p)^2 log\frac{m^2}{\Lambda^2} 
- 4 (v \cdot p)^2 F\left (\frac{m}{v \cdot p} \right ) + 2 (v \cdot p)^2
\ee
with 
\be
F(x) = \cases{  \sqrt{1 - x^2} \; tanh^{-1}(\sqrt{1 - x^2}) \qquad 0 \leq x \leq 1 \cr
                    \cr
              - \sqrt{x^2 -1 } \; tan^{-1}( \sqrt{x^2 -1}) \qquad x > 1 \cr }
\ee
\begin{eqnarray}
K_1(m) &=& \left [-m^2 + \frac{2}{3} (v \cdot p)^2 \right ] \, log \frac{m^2}{\Lambda^2} 
+ \frac{4}{3} \left [(v \cdot p)^2 - m^2 \right]  \nl
& * &  F\left (\frac{m}{ v\cdot p} \right ) -
 \frac{10}{9} ( v \cdot p)^2 + \frac{4}{3} m^2 - \frac{2 \pi}{3} \frac{m^3}{v \cdot p}, \nl
\end{eqnarray}
\be
R_x^{[3,2]}(m) = \frac{(m^2_{uu} - m^2) (m^2_{ss} - m^2)}{(m^2_{yy} - m^2)(m^2_\eta - m^2)},
\ee
\be
R_y^{[3,2]}(m) = \frac{(m^2_{uu} - m^2) (m^2_{ss} - m^2)}{(m^2_{xx} - m^2)(m^2_\eta - m^2)},
\ee
\be
R_\eta^{[3,2]}(m) = \frac{(m^2_{uu} - m^2) (m^2_{ss} - m^2)}{(m^2_{xx} - m^2)(m^2_{yy} - m^2)},
\ee
\begin{eqnarray}
R^{[2,2]}(m;{\cal I}) &=& \frac{(m^2_{uu} - m^2)(m^2_{ss} - m^2)}{(m^2_\eta - m^2)} 
{\cal I}(m) \nl
& +& \frac{(m^2_{uu} - m^2_\eta)(m^2_{ss} - m^2_\eta)}{(m^2 - m^2_\eta)} 
{\cal I}(m_\eta),  \nl
\end{eqnarray}
and
\be
DR^{[2,2]}(m;{\cal I}) = \frac{\partial}{\partial m^2} R^{[2,2]}(m; {\cal I}).
\ee

\vspace{.05in}
\noindent
We refer the reader to the original literature \cite{becirevic, bernard} 
for further details.  Here, for 
completeness, we give partially quenched 
formulas for the chiral logarithms in $D \rightarrow \pi$ 
 decays.  They will be used shortly in our own studies of $D \rightarrow \pi, l \nu$ 
decays. 
Some care is required in taking the $y \rightarrow x$ limit of (\ref{dfpara}), 
(\ref{dfperp}) and (\ref{dd}), however in the end expressions are simpler for 
$D \rightarrow \pi$ than for $D \rightarrow K$.
\begin{eqnarray}
 & & (4 \pi f)^2 \delta f^{D \rightarrow \pi}_\parallel = \nl
& & \left \{ (1 - 3 g^2_\pi) \, \left [ I_1(m_{xu}) + \frac{1}{2} I_1(m_{xs}) \right] 
+ \right . \nl
&& \left. \left [ 2 I_2(m_{xu}) + I_2(m_{xs}) \right ]
 - \frac{1+3g^2_\pi}{6} \left [ DR^{[2,2]}(m_{xx}; I_1) \right ] \right \} \nl
\end{eqnarray}
\begin{eqnarray}
 & & (4 \pi f)^2 \delta f^{D \rightarrow \pi}_\perp = \nl
& & \left \{ - (1 + 3 g^2_\pi) \, \left [ I_1(m_{xu}) + \frac{1}{2} I_1(m_{xs}) \right] 
+ \right . \nl
&  & \left. \frac{g^2_\pi}{3} \left [ DR^{[2,2]}(m_{xx}; K_1) \right ] 
- \frac{1 + 3g^2_\pi}{6} \left [ DR^{[2,2]}(m_{xx}; I_1) \right ] 
\right \} \nl
\end{eqnarray}
\begin{eqnarray}
& &  (4 \pi f)^2 \, D^{D \rightarrow \pi}  =  
 -3 g^2_\pi (v \cdot p) \; \times \nl
&& \left \{ \left [ 2 K_1(m_{xu}) + K_1(m_{xs}) 
\right ]  + 
\frac{1}{3} \left [ DR^{[2,2]}(m_{xx}; K_1) \right ] \right \} \nl
\end{eqnarray}




\end{document}